\documentclass[11pt,prd,showpacs,nofootinbib,oneside,hidelinks,a4paper,english]{revtex4-2}
\usepackage{verbatim}
\usepackage[pdftex]{pict2e}
\usepackage[dvipsnames]{xcolor}
\usepackage{tikz}
\usepackage{pgfplots}
\usepackage{feynmp-auto}
\usepackage{graphicx}
\usepackage{amssymb}
\usepackage{amsmath}
\usepackage[utf8]{inputenc}
\usepackage{textcomp}
\usepackage{dcolumn}
\usepackage{bm}
\usepackage{color}
\usepackage{epstopdf}
\usepackage{lipsum}
\usepackage{ulem}
\usepackage{braket}
\usepackage{hyperref}
\hypersetup{
	colorlinks=true,
	linkcolor=red,  
	citecolor=blue,  
	urlcolor=blue   
}

\begin{document}
	
	\title{Dissipative Multi-Field Dynamics from Non-Hermitian Inflationary Potentials}
	\author{S. D.  Campos}\email{sergiodc@ufscar.br}
	
	\affiliation{Applied Mathematics Laboratory-DFQM/CCTS, Federal University of São Carlos, Sorocaba, CEP 18052780, Brazil}
	
	\begin{abstract}
		In this work, we develop a perturbative framework for inflation driven by a complex inflaton with non-minimal gravitational coupling and a non-Hermitian potential. During the observable cosmic microwave background radiation era, the dynamics reduce to an effectively conservative two-field model, preserving the predictions of the $\alpha$-attractor class and satisfying Planck 2018 and BICEP/Keck constraints on $n_s$, $r$, and $f_{\mathrm{NL}}$. Near the end of inflation, trajectory bending activates the non-Hermitian sector, triggering geometric reheating. The resulting non-unitary evolution modifies the curvature spectrum and stochastic gravitational-wave background through a calculable damping factor determined by the complex mass eigenvalues. While cosmic microwave background-scale observables remain essentially unchanged, a distinctive suppression emerges in the high-frequency gravitational-wave spectrum ($f > 10^2$ Hz), potentially testable by future detectors such as the Einstein Telescope and the Big Bang Observer.
	\end{abstract}
	
	\maketitle
	
	\section{Introduction}
	
	The effective-field-theory (EFT) framework provides a powerful way to describe low-energy physics without requiring knowledge of its ultraviolet realization. By integrating out short-distance degrees of freedom, one obtains an effective potential that encodes the relevant dynamics of long-wavelength modes. This approach is particularly useful in inflationary cosmology, where the geometry of the scalar potential and its coupling to gravity determine the evolution of primordial perturbations and their observable signature in the cosmic microwave background (CMB) radiation and large-scale structure \cite{symanzik1970,demianski.1991,hertzberg2010,darabi2015,kodama2022,Aghanim:2018eyx}. 
	
	Among inflationary scenarios, $\alpha$-attractors are distinguished by their asymptotically flat potentials and robust predictions for the scalar spectral index $n_s$ and tensor-to-scalar ratio $r$ \cite{kallosh_2013_2,kallosh.2013_1,kallosh.2014,bhattacharya.2023,herrera2021,rodrigues2021,dimopoulos.2018}. Non-minimal couplings, such as $\xi\phi^2R$, naturally generate this plateau structure in the Einstein frame and suppress $r$, allowing models that are disfavored under minimal coupling to remain compatible with Planck and BICEP/Keck observations \cite{Aghanim:2018eyx,ade2021,demianski.1991,hertzberg2010,campos2026}.
	
	In previous work \cite{campos2026}, we showed that a complex inflaton field (CIF) \cite{khalatnikov.1992,khalatnikov.1994} with non-minimal coupling can yield an $\alpha$-attractor-like plateau through the real part of its potential, while its imaginary part induces an effective non-Hermitian deformation that becomes relevant near the end of inflation \cite{campos2026}. As the inflaton trajectory rotates in the internal field space $(x,\theta)$, this sector can trigger geometric reheating, transferring energy to an effective radiation bath and modifying the propagation of primordial perturbations and gravitational waves (GWs) \cite{campos2026}.
	
	The present work aims to develop a perturbative framework for the CIF model and determine the consequences of its non-Hermitian dynamics. During the observable slow-roll phase, the imaginary component is subdominant, and the system is effectively described by a conservative two-field theory in the Einstein frame. Using the adiabatic/entropy decomposition and the $\delta N$ formalism, we show that isocurvature modes and local non-Gaussianity are suppressed, leaving $n_s$, $r$, and $f_{\rm NL}^{\rm local}$ close to their single-field $\alpha$-attractor values.
	
	
	This paper is organized as follows. Section~\ref{sec:complex} introduces the complex quantities used throughout. Section~\ref{sec:perturbations} develops perturbation theory in the effectively Hermitian slow-roll regime. Section~\ref{sec:complex-quadratic} treats the genuinely non-Hermitian regime, while Section~\ref{sec:Pzeta-complex} derives the effective damping of curvature perturbations. Section~\ref{sec:pheno} discusses the resulting phenomenology, and Section~\ref{sec:conclusions} summarizes our conclusions.

	\section{Complex Inflaton Field and Complex Potential}\label{sec:complex}
	
	Complex fields arise in contemporary particle physics \cite{khalatnikov.1997} which naturally motivate their extension to inflationary scenarios \cite{campos2026}. A complex scalar field provides the most general quadratic EFT structure \cite{burgess.2007}, since the field and its conjugate can couple independently, leading to asymmetric and generally complex mass matrices that a single real scalar cannot yield. Here, the CIF is parameterized by $\Phi=\Phi(\phi,\chi)$ and its conjugate $\Phi^{*}=\Phi^{*}(\phi,\chi)$, which are defined in terms of the auxiliary real scalar fields $\phi$ and $\chi$ as
	\begin{equation}\label{eq:def1}
		\Phi=\frac{1}{\sqrt{2}}(\phi+i\chi), \qquad \Phi^{*}=\frac{1}{\sqrt{2}}(\phi-i\chi),
	\end{equation}
	and, consequently, $\Phi\Phi^{*}=|\Phi|^2=\frac{1}{2}(\phi^2+\chi^2)$. Observe that the CIF splits into a conservative sector governed by the real part of the potential and a non-conservative sector governed by the imaginary part. The latter is well suited to modeling instabilities, decay, and energy transfer in an expanding Universe \cite{burgess.2007,denner2006}. Phenomenologically, we treat the imaginary sector as a thermal reservoir, offering a coarse-grained description of reheating \cite{campos2026}.
	
	Current observations and theory do not point to a well-motivated EFT with a unique scalar potential that can empirically describe all cosmic epochs, from inflation and reheating through the hot big-bang era to today’s late-time acceleration \cite{guendelman2025,aguilar2023}. Instead, the usual framework combines several EFTs, each valid during a specific epoch. However, a promising route to a unified description is the class of $\alpha$-attractor models \cite{herrera2021,kallosh_2013_2,kallosh.2013_1,kallosh.2014}, which includes certain quintessential-inflation EFTs \cite{dimopoulos.2018,brissenden2024}. In these $\alpha$-quintessential scenarios, a single scalar degree of freedom, possibly with non-canonical kinetics or dissipative couplings \cite{basterogil.2021}, evolves on a multi-plateau potential: a high-energy inflationary plateau, a steep region that enables reheating and standard hot big-bang evolution, and a low-energy plateau or tail that drives today’s dark-energy acceleration \cite{dimopoulos.2018}. Concrete models of this type can be consistent with inflationary observables and late-time acceleration \cite{kallosh.2014}, and some also produce viable dark matter. Current data, however, do not yet favor a unique model within this class.
	
	For our purposes, one defines the complex potential $V(\Phi)$ through the real-valued functions $V_R$ and $V_I$ \cite{campos2026}
	\begin{equation}\label{eq:pot2}
		V(\Phi)=V_R(\Phi)+iV_I(\Phi),
	\end{equation}
	and, in particular, the potential for the real sector is constructed as a transition between a polynomial regime and an asymptotic plateau as \cite{campos2026}
	\begin{eqnarray}\label{eq:realpot}
		\nonumber V_R(\phi,\chi)&=&V_0\left[\frac{1}{2}(\varepsilon_\phi+\varepsilon_\chi)(\phi^2-\chi^2) -\frac{1}{2}m^2(\phi^2+\chi^2)+\frac{\lambda}{4}(\phi^2+\chi^2)^2\right]P(x)+\\&+& V_0\bigl[1-P(x)\bigr]=V_0{V_{p}P(x)+\bigl[1-P(x)\bigr]},
	\end{eqnarray}
	where the polynomial part $V_p=V_p(\phi,\chi)$ may, however, contain angular dependence through terms such as $\phi^2-\chi^2$. The radial plateau $P(x)$ is defined as a smoothed step function, similar to the $\alpha$-attractor inflation models mentioned previously, as
	\begin{eqnarray}\label{eq:plateau}
		P(x)=\frac{1}{1+(x/\mu)^s}
	\end{eqnarray}
	which acts as an EFT radial form factor, modifying the real part of the inflaton potential without introducing novel degrees of freedom. The parameter \(x \equiv \sqrt{\phi^{2} + \chi^{2}}\) measures the CIF amplitude and remains invariant under real orthogonal rotations in the \((\phi,\chi)\) plane whereas \(s\) controls the transition from the polynomial regime (\(x \ll \mu\)) to the asymptotic plateau (\(x \gg \mu\)). Typically, \(s=2\) gives a smooth transition, while \(s \gtrsim 6\) produces a sharp plateau; in what follows, we set \(s=2\). The quartic self-coupling \(\lambda\) is chosen to be small so that the potential remains sub-Planckian, \(V_R \ll M_P^4\), where $M_P$ stands for the Planck mass, and quantum gravity effects are negligible \cite{baumann.2012}. The mass parameter is also small, \(m = 10^{11}\)–\(10^{13}\,\text{GeV}\) (\(m \sim 10^{-8}\)–\(10^{-6}M_P\)) \cite{stein.2021}. Therefore, for definiteness, we assume \(\lambda = 10^{-13}\) and \(m = 10^{-6}M_P\), which keeps the inflaton potential below the Planck scale. Without loss of generality, the interaction couplings are settled so that the mean slope satisfies \((\varepsilon_\phi + \varepsilon_\chi)/2 = 1\). Taking these definitions into account, the model exhibits a radially controlled plateau in the large-field regime, with smooth angular anisotropies in the polynomial sector. 
	
	In general, the leading non-Hermitian operators are bilinear in the real components $(\phi,\chi)$ and can be written as 
	\begin{equation}\label{eq:pot_geral}
		V_I(\phi,\chi)
		= a_\phi\,\phi^2 + a_\chi\,\chi^2 + a_{\phi\chi}\,\phi\chi
		+ b_3\,\partial_\mu\phi\,\partial^\mu\chi + \Sigma \;,
	\end{equation}
	where $\Sigma$ denotes possible higher-derivative and higher-dimensional
	operators. At quadratic order, the imaginary sector may include diagonal terms $a_\phi\,\phi^2$, $a_\chi\,\chi^2$, and the off-diagonal mixing term $a_{\phi\chi}\,\phi\chi$. We adopt a minimal non-Hermitian deformation subject to two assumptions. First, the imaginary sector must remain negligible along the quasi-single-field slow-roll trajectories, so that CMB predictions stay continuously connected to the Hermitian $\alpha$-attractor limit. Diagonal imaginary mass terms would generally be active there and could affect both the amplitude and the spectral tilt of the scalar power spectrum in the observable range. Second, the non-Hermitian contribution should be triggered by angular motion in field space and should become relevant only when the trajectory bends near the end of inflation. 
	\begin{figure}[t]
		\centering
		\includegraphics[width=0.8\linewidth]{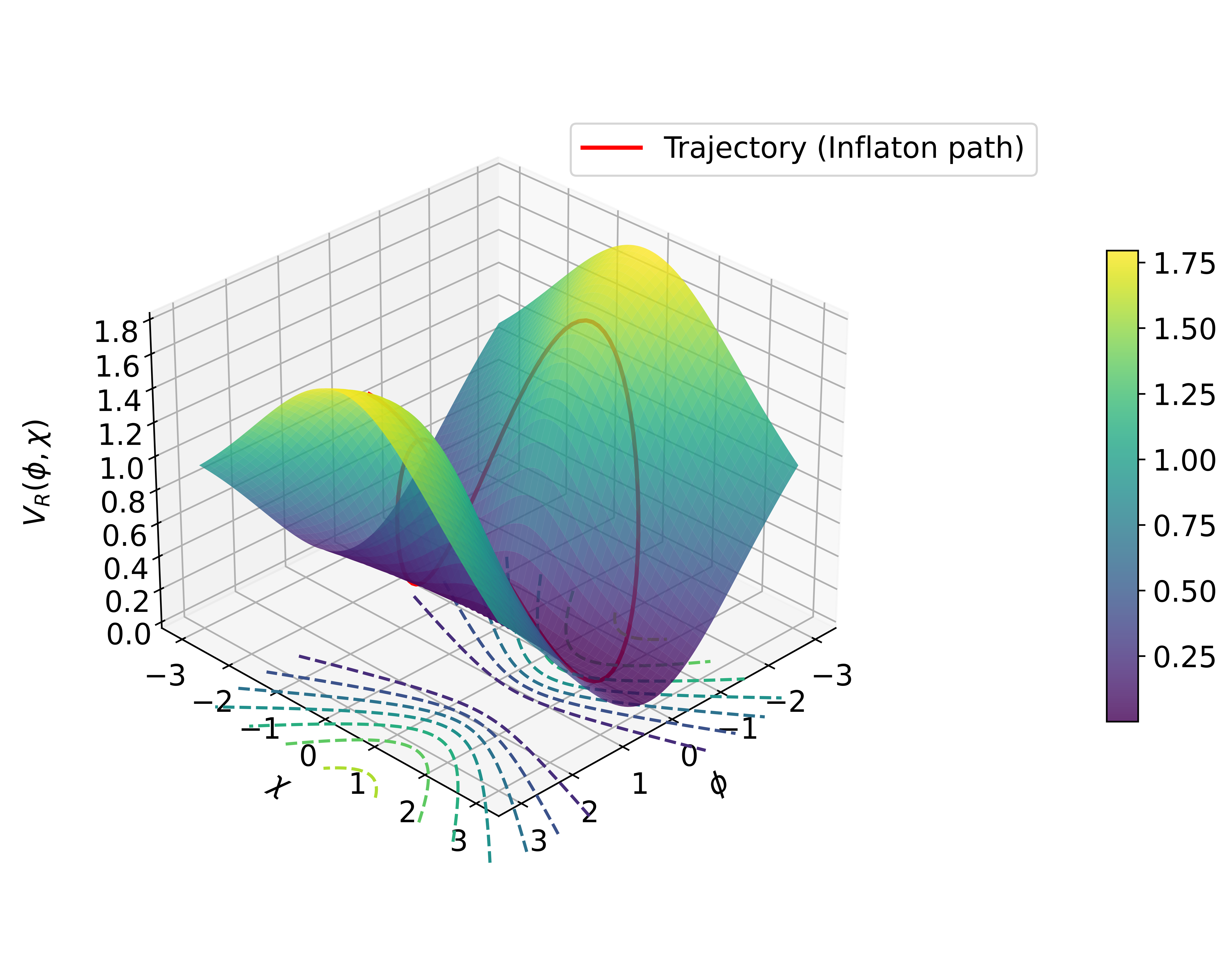}
		\vspace{-1.0cm}
		\caption{Three-dimensional plot of the real potential $V_R(\phi, \chi)$ with the background trajectory shown in red. The potential displays the typical plateau of $\alpha$-attractor models at large field values. The trajectory’s rotation near the minimum $(\phi, \chi) = (0,0)$ indicates activation of the non-Hermitian sector $V_I$, which drives geometric reheating and damps cosmological perturbations.}
		\label{fig:landscape}
	\end{figure}
	
	Taking the above assumptions into consideration, the lowest-dimensional operator that satisfies these conditions is the mixed bilinear term $\phi\chi$ \cite{campos2026}. Therefore, after setting the diagonal imaginary mass terms to zero, or equivalently treating them as subdominant in the minimal truncation, one assumes $a_{\phi\chi}=\varepsilon_\phi-\varepsilon_\chi = \Delta\varepsilon$ in equation \eqref{eq:pot_geral} and writes
	\begin{equation}
		V_I(\phi,\chi) = (\varepsilon_\phi - \varepsilon_\chi)\,\phi\,\chi
		\equiv \Delta\varepsilon\,\phi\,\chi \;,
	\end{equation}
	which is the leading non-Hermitian operator compatible with our field content and symmetries. Of course, $x$ in equation \eqref{eq:plateau} can be interpreted as the radial magnitude of the field in the corresponding internal field space. Writing the real components of the CIF $\phi=x\cos\theta$, $\chi=x\sin\theta$, the imaginary part of the potential becomes
	\begin{eqnarray}
		V_I(x,\theta)=\frac{\Delta\varepsilon}{2}x^2\sin 2\theta,
	\end{eqnarray}
	and hence the non‑Hermitian contribution is suppressed along particular directions in field space, $\theta=0,\pi/2$, and is controlled by the angular dynamics of $\theta(t)$. Therefore, within the present framework, inflation evolves approximately along a direction for which $V_I\simeq 0$, while a subsequent rotation in field space activates $V_I$, thereby providing a physical realization of delayed dissipative or $\mathcal{CP}$-violating effects at later stages.
	
	Figure~\ref{fig:landscape} shows the three-dimensional surface of the real potential $V_R(\phi, \chi)$ and the inflationary trajectory. The red solid line shows the background evolution, starting on the asymptotically flat plateau where $V_R \approx V_0$, and, near the end of inflation, the trajectory turns in the $(\phi, \chi)$ plane. This turn reflects a rotation in the effective state space that mixes the conservative dynamics with the dissipative term. As the trajectory turns, the model couples these two sectors so that the angular component is not only geometric. Actually, it acts as the bridge through which amplitude is transferred to the dissipative channel: the turning motion produces a nonzero coupling term between the conservative sector and $V_I$, so the angular change becomes the mechanism through which dissipation enters the dynamics. The dashed contours at the base show the underlying field-space metric $\mathcal{G}_{IJ}$ and the potential gradients that drive the field toward the minimum, setting the stage for the re-entry of perturbation modes and the consequent geometric reheating phase \cite{campos2026}.
	
	It is important to note that, to preserve the quantum consistency of the open system and satisfy the Fluctuation-Dissipation Theorem \cite{kamenev_book}, $V_I$ can be recast through a Hubbard-Stratonovich transformation. This converts the imaginary damping term into a Gaussian stochastic noise source $\xi(t, \mathbf{x})$ with correlator
	\begin{equation}
		\langle \xi(t,\mathbf{x})\,\xi(t',\mathbf{x}')\rangle = 2\,\mathcal{D}(\mathbf{x},\mathbf{x}')\,\delta(t-t')
	\end{equation}
	(or, in the local Markovian limit, $2D\,\delta(t-t')\,\delta^{(3)}(\mathbf{x}-\mathbf{x}')$), where the diffusion coefficient $D$ is intrinsically tied to $V_I$ and the bath temperature. Under the assumption of a Markovian bath, the noise is white in time, and its amplitude is constrained by the fluctuation-dissipation relation, thereby reformulating the field evolution into a stochastic Langevin equation.
	
	\section{Einstein-frame perturbations and non-Gaussianity}
	\label{sec:perturbations}
	
	
	
	Here, we establish a connection between the CIF model and large-scale cosmological observables, showing that non-Hermitian effects do not significantly alter the CMB physics.
	
	\subsection{Einstein-frame formulation and adiabatic/entropy basis}
	\label{subsec:EF-formulation}
	
	The effective Planck mass is determined by the real function $F(\phi, \chi)$, which accounts for the non-minimal coupling $\xi$
	\begin{equation}\label{eq:ftheta}
		F(\phi,\chi)=M_{\rm P}^2-\xi(\phi^2+\chi^2)>0,
	\end{equation}
	where $F(\phi,\chi) > 0$ ensures the stability of the gravitational interaction. In the Jordan frame, the resulting Klein–Gordon equations contain the non-minimal terms
	$-\xi R \phi$ and $-\xi R \chi$
	\begin{equation}\label{eq:eom_FRW}
		\begin{cases}
			\ddot\phi+3H\dot\phi-\zeta R\,\phi+\frac{\partial V}{\partial\phi}=0,\\
			\ddot\chi+3H\dot\chi-\zeta R\,\chi+\frac{\partial V}{\partial\chi}=0.
		\end{cases}
	\end{equation}
	where $R$ is the Ricci scalar curvature. To decouple the scalar fields from the Ricci scalar and work in the canonical Einstein frame, we perform a conformal transformation of the metric, $\tilde{g}_{\mu\nu} = \Omega^2(\phi,\chi) g_{\mu\nu}$, where the conformal factor is defined as
	\begin{equation}
		\Omega^2(\phi,\chi) = \frac{F(\phi,\chi)}{M_{\text{P}}^2} > 0.
	\end{equation}
	
	This field-dependent rescaling is chosen to normalize the coefficient of the Ricci scalar $\tilde{R}$ to $M_{\text{P}}^2/2$, ensuring that the gravitational sector assumes its standard Einstein form while preserving the metric signature. Then, one writes the action in the Einstein frame as
	\begin{equation}\label{eq:EF-action-CIF}
		S_{\rm E}
		\;=\;
		\int {\rm d}^4x \,\sqrt{-\tilde{g}}\,
		\left[
		\frac{M_{\rm P}^2}{2}\,\tilde{R}
		- \tilde{g}^{\mu\nu}\,
		\mathcal{G}\!\left(|\Phi|^2\right)\,
		\partial_\mu \Phi\,\partial_\nu \Phi^{*}
		- U(\Phi,\Phi^{*})
		\right],
	\end{equation}
	where $\mathcal{G}\!\left(|\Phi|^2\right)$ is the field-space metric induced by the conformal rescaling of the spacetime metric and by the Weyl transformation of the scalar fields (see Appendix \ref{app:field_space_metric}). The potential $V(\phi,\chi)$ in the Jordan frame is rescaled into the Einstein frame as
	\begin{equation}\label{eq:einstein_pot}
		U(\phi,\chi)=\frac{V(\phi,\chi)}{\Omega^4(\phi,\chi)}=\frac{V(\phi,\chi)M_P^4}{F^2(\phi,\chi)},
	\end{equation}
	and the scalar kinetic terms acquire a non-trivial $\mathcal{G}(|\Phi|^2)$. For convenience, we use the notations $U(\phi,\chi)$ and $U(\Phi,\Phi^{*})$ interchangeably, depending on the required representation.
	
	In the slow-roll regime relevant to CMB scales, we take $U(\Phi,\Phi^{*})$ to be the real part of the Jordan-frame potential after conformal rescaling, since the imaginary sector $\sim i\,\Delta\varepsilon\,\phi\chi$ is a small perturbation in this range, and its energy contribution is enclosed by the higher-order corrections \cite{campos2026}. Assuming an FRW background, $\mathrm{d}\tilde{s}^2 = -\mathrm{d}\tilde{t}^2 + a^2(\tilde{t})\mathrm{d}\mathbf{x}^2$, the Einstein-frame homogeneous equations read
	\begin{equation}\label{eq:EF-background-Friedmann}
		3 M_{\rm P}^2 \tilde{H}^2
		\;=\;
		\mathcal{G}\!\left(|\Phi|^2\right)\,\dot{\Phi}\,\dot{\Phi}^{*}
		+ U(\Phi,\Phi^{*}),
	\end{equation}
	\begin{equation}\label{eq:EF-background-KG}
		\tilde{D}_t \dot{\Phi}
		+ 3 \tilde{H}\,\dot{\Phi}
		+ \mathcal{G}\!\left(|\Phi|^2\right)\,U_{,\Phi^{*}} \;=\; 0,
		\qquad
		\tilde{D}_t \dot{\Phi}^{*}
		+ 3 \tilde{H}\,\dot{\Phi}^{*}
		+ \mathcal{G}\!\left(|\Phi|^2\right)\,U_{,\Phi} \;=\; 0,
	\end{equation}
	where overdots denote derivatives with respect to Einstein-frame cosmic time $\tilde{t}$, $U_{,\Phi} \equiv \partial U / \partial \Phi$, $U_{,\Phi^{*}} \equiv \partial U / \partial \Phi^{*}$, and
	$\tilde{D}_t$ denotes the usual covariant derivative in field space,
	\begin{equation}\label{eq:Dt-def-CIF}
		\tilde{D}_t \dot{\Phi}
		\;\equiv\;
		\ddot{\Phi} + \Gamma_{\Phi^{*}\Phi}^{\Phi}\,\dot{\Phi}^{*}\,\dot{\Phi},
		\qquad
		\tilde{D}_t \dot{\Phi}^{*}
		\;\equiv\;
		\ddot{\Phi}^{*} + \Gamma_{\Phi\Phi^{*}}^{\Phi^{*}}\,\dot{\Phi}\,\dot{\Phi}^{*},
	\end{equation}
	with $\Gamma_{\Phi^{*}\Phi}^{\Phi}$ and $\Gamma_{\Phi\Phi^{*}}^{\Phi^{*}}$ being the Christoffel symbols built from $\mathcal{G}\!\left(|\Phi|^2\right)$.
	Equations \eqref{eq:EF-background-Friedmann} and \eqref{eq:EF-background-KG} are the Einstein-frame counterparts of the Jordan-frame equations, reproducing the usual conservation law $\dot{\tilde{\rho}}+3\tilde{H}(\tilde{\rho}+\tilde{p})=0$. 
	
	Following the standard multi-field formalism \cite{Sasaki:1995aw,Gordon:2000hv,Wands:2007bd,Langlois:2008qf,Elliston:2011dr,Byrnes:2010em}, one determines the magnitude of the field velocity and the unit tangent vector along the background trajectory, respectively, as
	\begin{equation}\label{eq:sigma-def-CIF}
		\dot{\sigma}   \;\equiv\;\sqrt{2\,\mathcal{G}\!\left(|\Phi|^2\right)\,\dot{\Phi}\,\dot{\Phi}^{*}},
		\qquad
		e_\sigma^\Phi   \;\equiv\;   \frac{\dot{\Phi}}{\dot{\sigma}},
		\qquad   e_\sigma^{\Phi^{*}}  \;\equiv\;   \frac{\dot{\Phi}^{*}}{\dot{\sigma}},
	\end{equation}
	in which a unit vector $\{e_s^\Phi, e_s^{\Phi^{*}}\}$ orthogonal to $\{e_\sigma^\Phi, e_\sigma^{\Phi^{*}}\}$ with respect to $\mathcal{G}\!\left(|\Phi|^2\right)$
	is defined by
	\begin{equation}\label{eq:es-orthonormal-CIF}
		\mathcal{G}\!\left(|\Phi|^2\right)
		\left(e_\sigma^\Phi\,e_s^{\Phi^{*}} + e_\sigma^{\Phi^{*}}\,e_s^\Phi\right)
		= 0,
	\end{equation}
	with the trivial meaning that the adiabatic and entropy directions are orthogonal with respect to the field-space metric. Of course, the orthonormality of the kinematic basis is completed by 
	\begin{equation}\label{eq:es-orthonormal-CIF_2}
		\mathcal{G}\left(|\Phi|^2\right)
		\left(e_s^\Phi e_s^{\Phi^{}} + e_s^{\Phi^{}} e_s^\Phi\right)
		= 1,
	\end{equation}
	ensuring a consistent decomposition of the field perturbations. Consequently, the pairs $\{e_\sigma^\Phi,e_\sigma^{\Phi^{*}}\}$ and $\{e_s^\Phi,e_s^{\Phi^{*}}\}$ uniquely define the adiabatic and entropic directions, respectively, in the field-space manifold. By projecting the potential gradient along these directions, we obtain
	\begin{equation}\label{eq:Vproj-def-CIF}
		U_\sigma   \;\equiv\;   e_\sigma^\Phi\,U_{,\Phi^{*}} + e_\sigma^{\Phi^{*}}\,U_{,\Phi},
		\qquad
		U_s   \;\equiv\;   e_s^\Phi\,U_{,\Phi^{*}} + e_s^{\Phi^{*}}\,U_{,\Phi},
	\end{equation}
	and the resulting background equations can be written as
	\begin{equation}\label{eq:sigma-theta-eom-CIF}
		\ddot{\sigma} + 3 \tilde{H} \dot{\sigma} + U_\sigma = 0,
		\qquad
		\dot{\vartheta} = - \frac{U_s}{\dot{\sigma}},
	\end{equation}
	where $\vartheta$ is the angle that specifies the orientation of the basis vectors $\{e_\sigma^\Phi, e_\sigma^{\Phi^{*}}\}$, while $\omega \equiv \dot{\vartheta}$ represents the turning rate of the background trajectory in field space\footnote{In the approximately radial and weakly curved regime, the polar field-space angle \(\vartheta\) and the kinematic basis angle \(\theta\) may be identified to leading order. In the general curved-field-space case, however, they represent distinct quantities.}. This is a key result of the present perturbative analysis, since it allows a decomposition into the adiabatic mode, which governs curvature perturbations, and the entropy mode, which encodes isocurvature fluctuations and describes how energy is redistributed among field degrees of freedom during inflation.
	
	\subsection{Linear adiabatic and entropy perturbations}
	\label{subsec:linear-perturbations-CIF}
	
	Of course, by adopting the spatially flat gauge, the spatial curvature perturbation is set to zero, and the cosmological information is entirely encoded in the scalar field fluctuations $\delta\Phi$ and $\delta\Phi^{*}$, projected onto the kinematic basis to yield the adiabatic $Q_\sigma=Q_\sigma(k,\tilde{t})$ and entropy $Q_s=Q_s(k,\tilde{t})$ modes
	\begin{equation}
		Q_\sigma   \;\equiv\;   e_\sigma^\Phi \delta\Phi + e_\sigma^{\Phi^{*}} \delta\Phi^{*},
		\qquad
		Q_s   \;\equiv\;   e_s^\Phi \delta\Phi + e_s^{\Phi^{*}} \delta\Phi^{*},
		\label{eq:Qsigma-Qs-def-CIF}
	\end{equation}
	where the time \(\tilde{t}\) is measure after Hubble exit. To linear order, the Fourier modes $Q_\sigma$ and
	$Q_s$ obey the coupled equations \cite{Gordon:2000hv,Langlois:2008qf}
	\begin{align}
		\ddot{Q}_\sigma
		&+ 3 \tilde{H} \dot{Q}_\sigma
		+ \left[
		\frac{k^2}{a^2}
		+ U_{\sigma\sigma}
		- \omega^2
		- \frac{1}{a^3}
		\frac{{\rm d}}{{\rm d}\tilde{t}}
		\left(
		\frac{a^3 \dot{\sigma}^2}{\tilde{H}}
		\right)
		\right] Q_\sigma
		\;=\;
		2 \omega \dot{Q}_s,
		\label{eq:Qsigma-eq-CIF}
		\\
		\ddot{Q}_s
		&+ 3 \tilde{H} \dot{Q}_s
		+ \left[
		\frac{k^2}{a^2}
		+ U_{ss}
		+ 3 \omega^2
		\right] Q_s
		\;=\; 0,
		\label{eq:Qs-eq-CIF}
	\end{align}
	where $k$ is the comoving wavenumber and 
	\begin{eqnarray}
		\nonumber U_{\sigma\sigma} &\equiv 
		e_\sigma^\Phi e_\sigma^{\Phi^{*}} \nabla_\Phi \nabla_{\Phi^{*}} U
		+ e_\sigma^{\Phi^{*}} e_\sigma^\Phi \nabla_{\Phi^{*}} \nabla_\Phi U, 
		\qquad
		\nonumber U_{ss}  &\equiv
		e_s^\Phi e_s^{\Phi^{*}} \nabla_\Phi \nabla_{\Phi^{*}} U
		+ e_s^{\Phi^{*}} e_s^\Phi \nabla_{\Phi^{*}} \nabla_\Phi U,
		\label{eq:Uss-def-CIF}
	\end{eqnarray}
	represent the second covariant derivatives of the potential projected along the adiabatic and entropy directions. Moreover, $\nabla_\Phi$ and $\nabla_{\Phi^{*}}$ are the covariant derivatives associated with the field-space metric $\mathcal{G}\left(|\Phi|^2\right)$.
	
	Deep inside the horizon, $k \gg a \tilde{H}$, allowing us to impose the Bunch–Davies initial conditions for the canonically normalized modes $Q_\sigma$ and $Q_s$
	\begin{equation}
		Q_a^k(\tau_e) = \frac{1}{\sqrt{2k}}e^{-ik \tau_e},\quad \partial _\tau Q_a^k(\tau_e) = -\frac{i}{\sqrt{2/k}}e^{-ik \tau_e},
	\end{equation}
	for $a\in\{\sigma,s\}$. In terms of the adiabatic mode, the comoving curvature perturbation $\zeta$ is given by \cite{Gordon:2000hv}
	\begin{equation}
		\zeta(k)
		\;=\;
		\frac{\tilde{H}}{\dot{\sigma}}\; Q_\sigma(k),
		\label{eq:zeta-from-Qsigma}
	\end{equation}
	showing that the adiabatic field fluctuations along the inflaton trajectory are directly responsible for the curvature (density) perturbations. 
	
	
	Once the modes are super-horizon and have frozen, the dimensionless power spectra for the curvature and entropy perturbations can be expressed as \cite{malik.2009, Gordon:2000hv, Langlois:2008qf}
	\begin{equation} \label{eq:power_spectra}
		P_\zeta(k) = \frac{k^3}{2\pi^2} \left| \frac{\tilde{H}}{\dot{\sigma}} Q_\sigma(k) \right|^2, \qquad P_s(k) \propto \frac{k^3}{2\pi^2} |Q_s(k)|^2,
	\end{equation}
	where the proportionality factor for $P_s$ depends on the specific normalization convention adopted for the entropy field $S(k)$ relative to the physical fluctuation $Q_s(k)$. By evaluating these spectra at the end of inflation, we can determine the isocurvature fraction and verify its consistency against CMB observations. 
	
	Solving (numerically) equations~\eqref{eq:Qsigma-eq-CIF} and \eqref{eq:Qs-eq-CIF} along the background trajectory allows us to compute the curvature and isocurvature power spectra, $P_\zeta(k)$ and $P_s(k)$, and, in particular, the isocurvature fraction at the end of inflation, written here as
	\begin{equation}
		\beta_{\rm iso}(k)
		\;\equiv\;
		\frac{P_s(k)}{P_\zeta(k) + P_s(k)}.
		\label{eq:beta-iso-def-CIF}
	\end{equation}
	
	The case $\beta_{\rm iso}(k)\approx 0$ means that nearly all scalar power at that scale is adiabatic: the initial conditions are effectively single-field–like, as in standard $\alpha$-attractor inflation. Conversely, when $\beta_{\rm iso}(k)\approx 1$, the adiabatic perturbations are subdominant. Therefore, in the parameter region where $\Delta\varepsilon$ is small and the imaginary sector acts as a spectator during slow roll, we find that the turning rate $\omega$ and the entropy mode $Q_s$ remain strongly suppressed at CMB scales, leading to a negligible isocurvature fraction and a curvature spectrum that is effectively single-field and $\alpha$-attractor–like, in agreement with our Einstein-frame mapping of the background \cite{campos2026}.

	\subsection{Super-horizon evolution and \texorpdfstring{$\delta N$}{delta N}}
	\label{subsec:deltaN-CIF}
	
	On super-horizon scales, the curvature perturbation $\zeta$ equals the perturbation in the number of e-folds between an initial spatially flat and a final uniform-density hypersurface \cite{Maggiore:2018sht}
	\begin{equation}
		N = \ln\left(\frac{a_{\text{final}}}{a_{\text{initial}}}\right),
	\end{equation}
	where $N$ is the number of e-folds. Within the $\delta N$ formalism \cite{starobinsky.1985,Sasaki:1995aw,lyth.2005}, one writes
	\begin{equation}
		\zeta(\mathbf{x})   \;=\;   \delta N(\mathbf{x})
		\;=\;   N\bigl[\Phi_\star + \delta \Phi_\star(\mathbf{x}), \Phi^{*}_\star + \delta \Phi^{*}_\star(\mathbf{x})\bigr]
		- N(\Phi_\star, \Phi^{*}_\star),
		\label{eq:deltaN-def-CIF}
	\end{equation}
	where $\Phi_\star,\Phi_\star^{*}$ denotes the background
	field evaluated on the initial flat slice, typically chosen a few
	e-folds after the Hubble exit of the pivot mode; $N(\Phi_\star,\Phi_\star^{*})$ is the homogeneous number of e-folds up to the final uniform-density slice. Expanding $\delta N$ to second order in the field perturbations yields (omitting $x$, for shortness)
	\begin{equation}
		\zeta
		\;=\;
		\delta N
		\;\simeq\;
		N_\Phi\,\delta \Phi_\star + N_{\Phi^{*}}\,\delta \Phi^{*}_\star
		+ \frac{1}{2} N_{\Phi\Phi}\,\delta \Phi_\star \delta \Phi_\star
		+ N_{\Phi\Phi^{*}}\,\delta \Phi_\star \delta \Phi^{*}_\star
		+ \frac{1}{2} N_{\Phi^{*}\Phi^{*}}\,\delta \Phi^{*}_\star \delta \Phi^{*}_\star
		+ \cdots,
		\label{eq:deltaN-expansion-CIF}
	\end{equation}
	where
	\begin{eqnarray}
		\nonumber N_\Phi\equiv \frac{\partial N}{\partial \Phi_\star},\quad
		N_{\Phi^{*}}\equiv \frac{\partial N}{\partial \Phi^{*}_\star},\quad
		N_{\Phi\Phi}\equiv \frac{\partial^2 N}{\partial \Phi_\star \partial \Phi_\star},\quad
		N_{\Phi\Phi^{*}}\equiv \frac{\partial^2 N}{\partial \Phi_\star \partial \Phi^{*}_\star},\quad
		N_{\Phi^{*}\Phi^{*}}\equiv \frac{\partial^2 N}{\partial \Phi^{*}_\star \partial \Phi^{*}_\star}.
		\label{eq:N-derivs-def-CIF}
	\end{eqnarray}
	are evaluated along the unperturbed Einstein-frame trajectory. In the CIF setup, we compute $N_\Phi$, $N_{\Phi^{*}}$, and their second derivatives numerically by integrating the background system
	\eqref{eq:EF-background-Friedmann}-\eqref{eq:EF-background-KG} over a grid of initial conditions
	$(\Phi_\star, \Phi^{*}_\star)$ near the best-fit solution and finite-differencing the resulting function $N(\Phi_\star, \Phi^{*}_\star)$. Assuming that the field perturbations at Hubble exit are nearly Gaussian and characterized by the covariance \cite{Sasaki:1995aw,nibbelink.2002}
	\begin{equation}
		\bigl\langle
		\delta \Phi_\star(\mathbf{k})\,
		\delta \Phi^{*}_\star(\mathbf{k}')
		\bigr\rangle
		\;=\;
		(2\pi)^3 \delta^{(3)}(\mathbf{k} + \mathbf{k}')
		\left(\frac{H_\star}{2\pi}\right)^2
		\mathcal{G}^{-1}(|\Phi_\star|^2),
		\label{eq:field-covariance-CIF}
	\end{equation}
	then the power spectrum of $\zeta$ is given by
	\begin{equation}
		P_\zeta(k)
		\;=\;
		2\left(\frac{H_\star}{2\pi}\right)^2
		N_\Phi N_{\Phi^{*}} \mathcal{G}^{-1}(|\Phi_\star|^2),
		\label{eq:Pzeta-deltaN-CIF}
	\end{equation}
	evaluated at the time when mode $k$ crosses the Hubble radius during inflation. While equation \eqref{eq:power_spectra} describes the full dynamical evolution of the power spectrum, encoding non-Hermitian dissipative effects, the $\delta N$ formalism presents an alternative for evaluating the spectrum at horizon exit, given by equation \eqref{eq:Pzeta-deltaN-CIF}, which allows a direct calculation of the spectral index and non-Gaussianity parameters.
	
	The bispectrum of $\zeta$ in the squeezed (local) configuration can be expressed in terms of the same derivatives, and the corresponding local-type non-linearity parameter is \cite{lyth.2005,vernizzi.2006,Byrnes:2010em,sasaki.2006}
	\begin{equation}
		f_{\rm NL}^{\rm local}
		\;\simeq\;
		\frac{5}{6}\left\{
		\frac{N_\Phi N_{\Phi^{*}} \left(N_{\Phi\Phi} + N_{\Phi^{*}\Phi^{*}}\right) + N_\Phi^2 N_{\Phi\Phi^{*}} + N_{\Phi^{*}}^2 N_{\Phi\Phi^{*}}}
		{\bigl[2 N_\Phi N_{\Phi^{*}} \mathcal{G}^{-1}(|\Phi_\star|^2)\bigr]^2},
		\label{eq:fNL-local-CIF}\right\}
	\end{equation}
	where the indices are raised and lowered using the field-space metric $\mathcal{G}(|\Phi|^2)$. In practice, we evaluate equation~\eqref{eq:fNL-local-CIF} numerically by finite-differencing $N(\Phi_\star, \Phi^{*}_\star)$ in a small neighborhood of the background trajectory in the $(\phi,\chi)$ plane.
	
	Notice that in the region of parameter space where the Einstein-frame background is well described by an effective single-field $\alpha$-attractor and the entropy mode is heavy or only weakly excited at Hubble exit, the mixed second derivatives in the Hessian are strongly suppressed relative to the product of first derivatives. As a result, the local-type non-Gaussianity from the CIF remains small, $|f_{\rm NL}^{\rm local}| \ll 1$, consistent with effectively single-field behavior on CMB scales. Significant deviations, larger $|f_{\rm NL}^{\rm local}|$, are expected only in narrow regions of the parameter space where the trajectory bends strongly in field space or $V_I$ becomes dynamically relevant at horizon crossing. 
	\begin{figure}[t]
		\centering
		\includegraphics[width=0.8\linewidth]{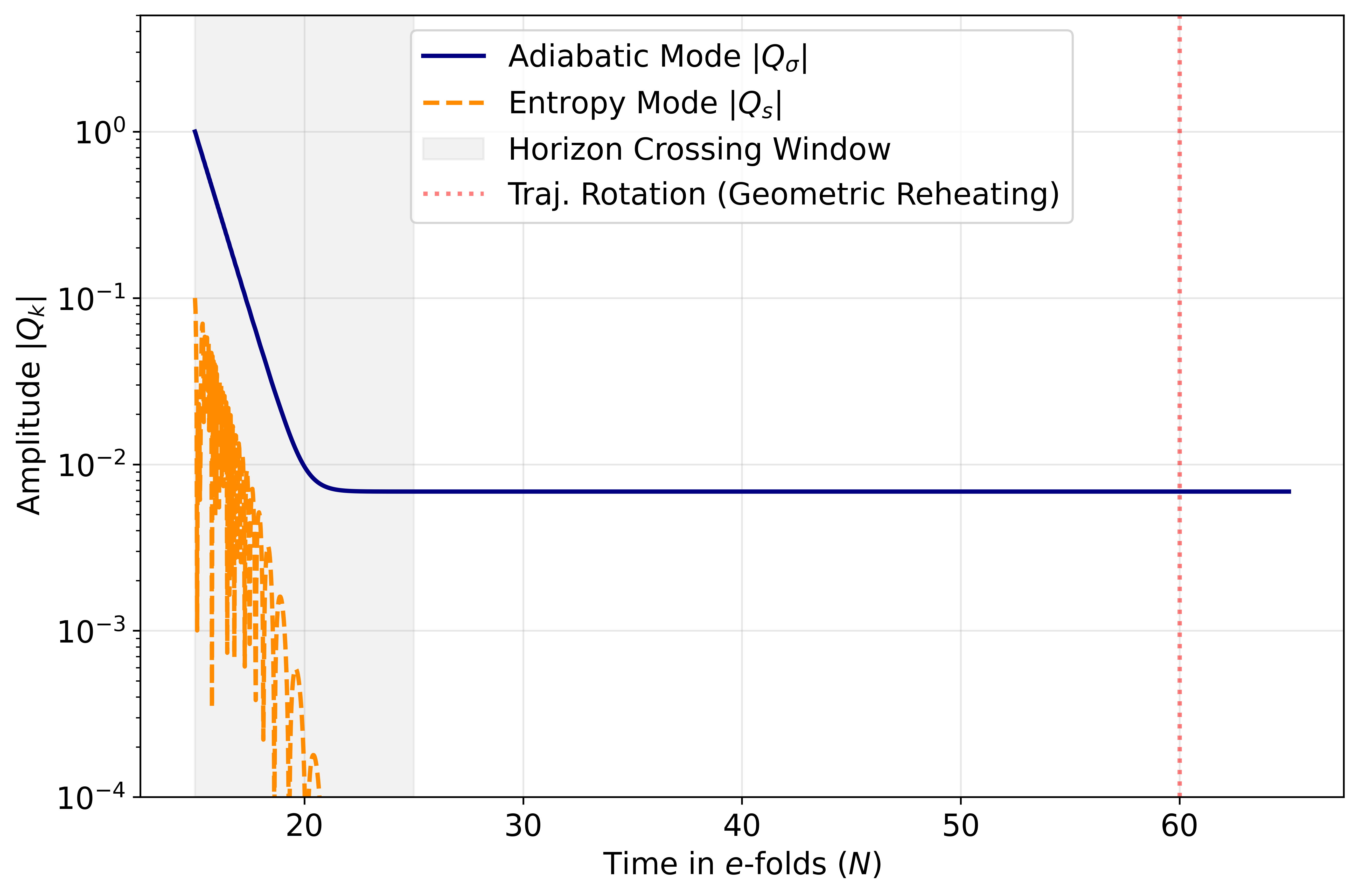}
		\caption{Consistent evolution of adiabatic ($|Q_{\sigma}|$) and entropy ($|Q_{s}|$) scalar perturbations. The gray band marks the horizon-crossing window ($N \approx 15\text{–}25$), where modes leave the Hubble radius. The adiabatic mode (solid blue) freezes at a constant amplitude on super-horizon scales, whereas the entropy mode (dashed orange) oscillates rapidly and decays, thereby suppressing isocurvature perturbations. The vertical red dotted line at $N=60$ marks the onset of geometric reheating, well after CMB scales have stabilized.}
		\label{fig:consistent_evolution}
	\end{figure}
	
	Figure \ref{fig:consistent_evolution} shows that the CIF is consistent with standard inflation by illustrating how scalar modes evolve across horizon crossing. Notice that an important requirement for a viable inflationary model is that curvature perturbations quickly settle into a constant value after leaving the Hubble radius \cite{lyth.2005jcap}. Observe that in the shaded region ($15 \lesssim N \lesssim 25$), the adiabatic mode amplitude $|Q_{\sigma}|$ shifts from sub-horizon evolution to a stable super-horizon plateau, allowing CMB temperature anisotropies. At the same time, the entropy mode $|Q_s|$ undergoes strongly damped oscillations and decreases by several orders of magnitude as it exits the horizon. This rapid decay suppresses isocurvature fluctuations, making the model effectively single-field on large scales, in agreement with Planck constraints \cite{Aghanim:2018eyx}. Furthermore, this figure shows a scale separation between the stabilization of primordial fluctuations and the onset of the non-Hermitian phase. The vertical line at $N=60$ marks the onset of trajectory rotation and geometric reheating, while the adiabatic amplitude remains constant well before this phase, indicating that the non-Hermitian potential deformation, though crucial for reheating and GW production, does not compromise the scalar spectral predictions. 
	
\begin{figure}[t]
		\centering
		\includegraphics[width=0.8\linewidth]{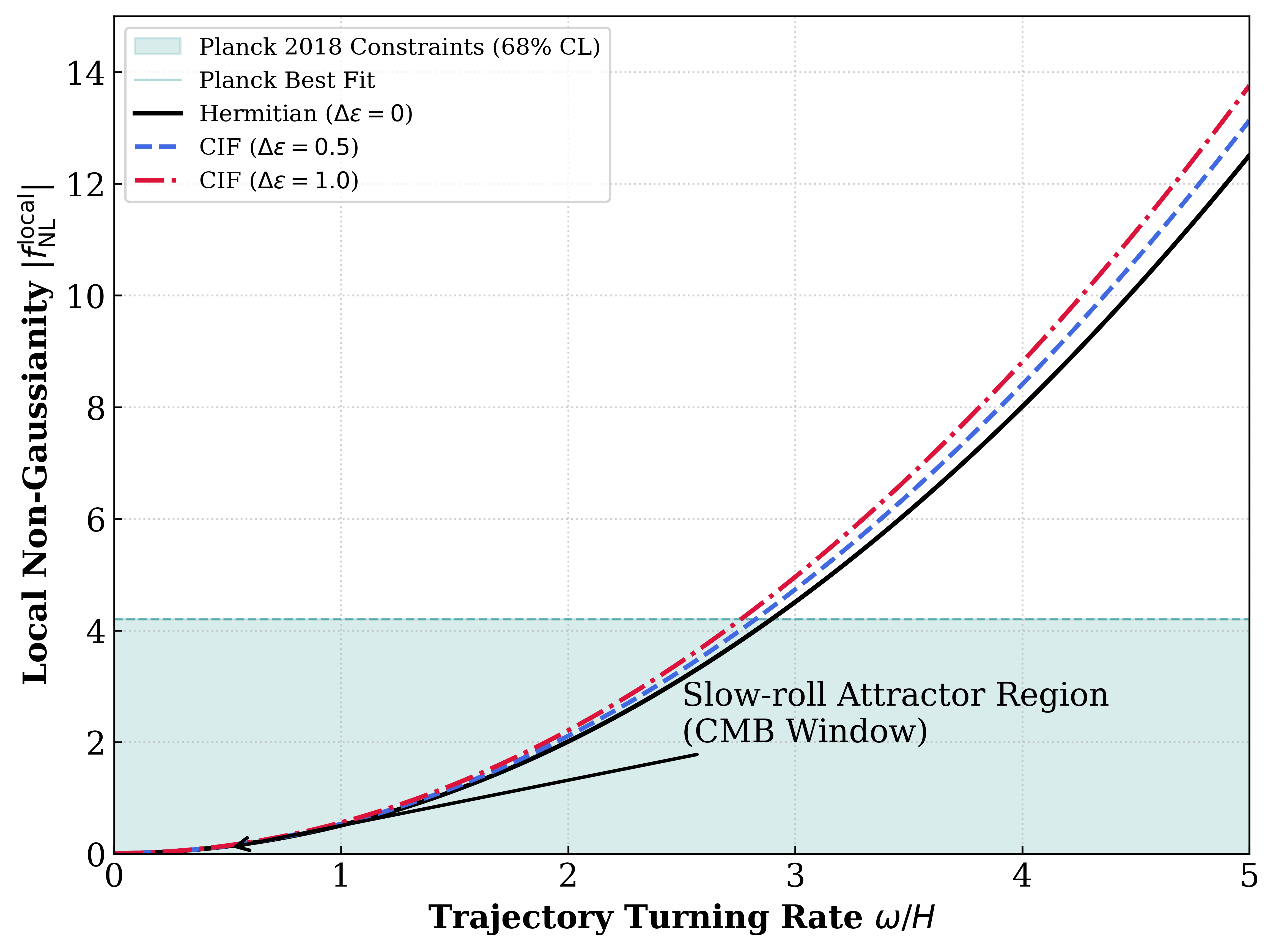}
		\caption{Local-type non-Gaussianity $|f_{\text{NL}}^{\text{local}}|$ as a function of trajectory turning rate $\omega/H$ for the Hermitian case ($\Delta\epsilon=0$) and non-Hermitian deformations ($\Delta\epsilon=0.5, 1.0$). The teal band indicates the $1\sigma$ constraints from Planck 2018. The "Slow-roll Attractor Region" marks where the model agrees with CMB data, showing that the non-Hermitian sector provides only a subdominant correction to the geometry-induced non-Gaussianity.}
		\label{fig:non_gaussianity}
	\end{figure}

	Figure \ref{fig:non_gaussianity} compares the CIF model with current observational bounds on primordial non-Gaussianity. Using the $\delta N$ formalism, we calculate the local-type non-linearity parameter $f_{\text{NL}}^{\text{local}}$ as a function of the turning rate $\omega/H$, which measures departures from a straight path in field space. In the "slow-roll attractor region" \cite{Linde1990,SalopekBond1990}, where the turning rate is low ($\omega/H \lesssim 2$), the predicted non-Gaussianity remains within the $1\sigma$ Planck 2018 bounds \cite{Aghanim:2019bto}. In this regime, the inflaton effectively behaves as a single field, and isocurvature contributions to curvature perturbations are strongly suppressed.
	
\section{Complex quadratic action and non-unitary mode equations}
\label{sec:complex-quadratic}

Here, the CIF framework is extended to cosmological perturbations, showing that $V_I$ produces a complex quadratic action for scalar fluctuations and non-unitary mode evolution, remaining consistent with $\mathcal{PT}$-symmetry. 


Following the standard perturbative approach, the CIF can be expressed as a linear combination of a time-dependent background $\bar{\Phi}(t)$ ($\bar{\Phi}^{*}(t)$) and small fluctuations $\delta\Phi(\mathbf{x},t)$ ($\delta\bar{\Phi}^{*}(\mathbf{x},t)$) as
\begin{equation}
	\Phi(\mathbf{x},t) = \bar{\Phi}(t) + \delta\Phi(\mathbf{x},t),
	\qquad
	\Phi^{*}(\mathbf{x},t) = \bar{\Phi}^{*}(t) + \delta\Phi^{*}(\mathbf{x},t).
\end{equation}
allowing us to expand the action given by equation~\eqref{eq:EF-action-CIF} to second order in $\delta\Phi$ and $\delta\Phi^{*}$. Transforming into Fourier space through $\delta\Phi(t,\mathbf{x}) \to \delta\Phi_{\mathbf{k}}(t)$ yields the quadratic action
\begin{eqnarray}
	\nonumber   S^{(2)} = \frac{1}{2} \int {\rm d}t\,{\rm d}^3k\,a^3
	\left[\mathcal{K}_{\Phi\Phi^{*}}\,\delta\dot{\Phi}_{\mathbf{k}}\,\delta\dot{\Phi}^{*}_{-\mathbf{k}}+ \mathcal{K}_{\Phi^{*}\Phi}\,\delta\dot{\Phi}^{*}_{\mathbf{k}}\,
	\delta\dot{\Phi}_{-\mathbf{k}} + \right.\\ 
	\left. -\delta\Phi_{\mathbf{k}}\,\mathcal{M}^2_{\Phi\Phi^{*}}\,\delta\Phi^{*}_{-\mathbf{k}} - \delta\Phi^{*}_{\mathbf{k}}\,
	\mathcal{M}^2_{\Phi^{*}\Phi}\,\delta\Phi_{-\mathbf{k}} + \cdots \right],
	\label{eq:quadratic-action-complex}
\end{eqnarray}
where $\mathcal{K}_{\Phi\Phi^{*}}$ and $\mathcal{K}_{\Phi^{*}\Phi}$ are real kinetic coefficients given by the field-space metric evaluated on the background, $\mathcal{K}{\Phi\Phi^{*}} = \mathcal{K}{\Phi^{*}\Phi} = \mathcal{G}(|\bar{\Phi}|^2)$. Of course, the complex mass matrix components $\mathcal{M}^2_{\Phi\Phi^{*}}$ and $\mathcal{M}^2_{\Phi^{*}\Phi}$ are given by
\begin{equation}
	\mathcal{M}^2_{\Phi\Phi^{*}}   =   \nabla_\Phi \nabla_{\Phi^{*}}U(\bar{\Phi},\bar{\Phi}^{*}) + \text{(curvature and non-minimal terms)}.
\end{equation}

It should be noted that the non-Hermitian quadratic action in equation \eqref{eq:quadratic-action-complex} is understood as a semi-classical effective description of an open quantum system. In a complete microscopic treatment, this dissipative action describes the evolution of the reduced density matrix after integrating out high-frequency environmental modes within the Schwinger-Keldysh formalism \cite{kamenev_book,Haehl2017}.

Terms depending on the spacetime coordinate can be eliminated by working in the spatially flat gauge, in which the spatial curvature perturbation and the shear perturbation are set to zero \cite{Gordon:2000hv,maldacena.2003,malik.2009}. In this gauge, the quadratic action reduces to a canonical form involving only the kinetic and mass matrices for the field perturbations, leading to the mode equations
\begin{equation}
	\ddot{\delta\Phi}_{\mathbf{k}} + 3\tilde{H}\,\dot{\delta\Phi}_{\mathbf{k}} + \left(\frac{k^2}{a^2} + \mathcal{M}^{2}_{\Phi\Phi^{*}}\right)\delta\Phi^{*}_{\mathbf{k}}= 0,
	\qquad
	\ddot{\delta\Phi}^{*}_{\mathbf{k}}
	+ 3\tilde{H}\,\dot{\delta\Phi}^{*}_{\mathbf{k}}+ \left(\frac{k^2}{a^2}
	+ \mathcal{M}^{2}_{\Phi^{*}\Phi}\right)\delta\Phi_{\mathbf{k}}= 0,
	\label{eq:mode-eq-complex}
\end{equation}
where the Hessian of \(U\) preserves the complex structure induced by the original potential, admitting the decomposition
\begin{eqnarray}
	\mathcal{M}^2_{\Phi\Phi^{*}} \;=\; \mathcal{M}^{2\,R}_{\Phi\Phi^{*}} \;+\; i\,\mathcal{M}^{2\,I}_{\Phi\Phi^{*}},
\end{eqnarray}
where (\(\mathcal{M}^{2\,R}_{\Phi\Phi^{*}}, \mathcal{M}^{2\,I}_{\Phi\Phi^{*}})\in \mathbb{R}\). Observe that the equation \eqref{eq:mode-eq-complex} generalizes the two-field equations derived for the case of a complex potential \cite{campos2026}.


From equation~\eqref{eq:sigma-def-CIF}, and respecting the orthonormality conditions in equations~\eqref{eq:es-orthonormal-CIF} and \eqref{eq:es-orthonormal-CIF_2}, the complexified adiabatic and entropic perturbations are given by
\begin{equation}
	Q_\sigma \equiv e_\sigma^\Phi \,\delta\Phi + e_\sigma^{\Phi^{*}} \,\delta\Phi^{*},
	\qquad
	Q_s \equiv e_s^\Phi \,\delta\Phi + e_s^{\Phi^{*}} \,\delta\Phi^{*},
\end{equation}
and projecting \eqref{eq:mode-eq-complex} along these directions, and including the usual connection terms associated with $\omega$, we obtain the following coupled equations
\begin{align}
	\ddot{Q}_\sigma &+ 3\tilde{H}\,\dot{Q}_\sigma+ \left[\frac{k^2}{a^2} + \mathcal{M}_{\sigma\sigma}^2- \omega^2- \frac{1}{a^3}\frac{{\rm d}}{{\rm d}\tilde{t}}\left(\frac{a^3 \dot{\sigma}^2}{\tilde{H}}\right)\right] Q_\sigma=2\omega\,\dot{Q}_s+ \mathcal{M}_{\sigma s}^2 Q_s,
	\label{eq:Qsigma-complex}
	\\
	\ddot{Q}_s&+ 3\tilde{H}\,\dot{Q}_s+ \left[\frac{k^2}{a^2}+ \mathcal{M}_{ss}^2+ 3\omega^2\right] Q_s=- 2\omega\,\dot{Q}_\sigma+ \mathcal{M}_{s\sigma}^2 Q_\sigma,
	\label{eq:Qs-complex}
\end{align}
where
\begin{align}
	\mathcal{M}_{\sigma\sigma}^2 &\equiv\left(e_\sigma^\Phi e_\sigma^{\Phi^{*}} + e_\sigma^{\Phi^{*}} e_\sigma^\Phi\right) \mathcal{M}^2_{\Phi\Phi^{*}},
	\quad
	\mathcal{M}_{ss}^2\equiv\left(e_s^\Phi e_s^{\Phi^{*}} + e_s^{\Phi^{*}} e_s^\Phi\right) \mathcal{M}^2_{\Phi\Phi^{*}},
	\\
	\mathcal{M}_{\sigma s}^2 &\equiv\left(e_\sigma^\Phi e_s^{\Phi^{*}} + e_\sigma^{\Phi^{*}} e_s^\Phi\right) \mathcal{M}^2_{\Phi\Phi^{*}},
	\quad
	\mathcal{M}_{s\sigma}^2\equiv\left(e_s^\Phi e_\sigma^{\Phi^{*}} + e_s^{\Phi^{*}} e_\sigma^\Phi\right) \mathcal{M}^2_{\Phi\Phi^{*}}.
\end{align}

When the imaginary sector is negligible, one has the obvious result $\mathcal{M}^2_{\Phi\Phi^{*}} \to \mathcal{M}^{2\,R}_{\Phi\Phi^{*}}$, and equations~\eqref{eq:Qsigma-complex} and \eqref{eq:Qs-complex} reduce to the standard real two-field system. However, when $V_I$ grows large near the end of inflation, the projected mass eigenvalues acquire imaginary components, and the perturbation evolution becomes non-unitary, modeling dissipative energy transfer from the inflaton to the reheating bath while remaining consistent with the background dynamics.


\section{Curvature power spectrum with complex mass eigenvalues}
\label{sec:Pzeta-complex}

Here, we derive a compact expression for the curvature power spectrum when the quadratic mass matrix of the complex inflaton is non-null. Using the complexified adiabatic/entropy equations \eqref{eq:Qsigma-complex}–\eqref{eq:Qs-complex}, the spectrum equals the standard Hermitian result multiplied by a damping factor determined by the imaginary parts of the eigenfrequencies of the complex projected masses.

\subsection{Mode diagonalization and complex frequencies}

For a given comoving wavenumber \(k\), it is convenient to write the coupled system \eqref{eq:Qsigma-complex}–\eqref{eq:Qs-complex} as
\begin{equation}
	\ddot{\mathbf{Q}}_{\mathbf{k}} + 3\tilde{H}\,\dot{\mathbf{Q}}_{\mathbf{k}}
	+ \mathbf{\Omega}^2_{\mathbf{k}}\,\mathbf{Q}_{\mathbf{k}} = 0,
	\qquad
	\mathbf{Q}_{\mathbf{k}} =
	\begin{pmatrix}
		Q_\sigma(\mathbf{k}) \\
		Q_s(\mathbf{k})
	\end{pmatrix},
	\label{eq:mode-matrix}
\end{equation}
where the complex frequency matrix \(\mathbf{\Omega}^2_{\mathbf{k}}\) is
defined by
\begin{equation}\label{eq:mat_eigenx}
	\mathbf{\Omega}^2_{\mathbf{k}}   =
	\begin{pmatrix}
		k^2/a^2 + \mathcal{M}_{\sigma\sigma}^2
		- \omega^2- \displaystyle\frac{1}{a^3}\frac{{\rm d}}{{\rm d}\tilde{t}}
		\Bigl(\frac{a^3\dot{\sigma}^2}{\tilde{H}}\Bigr)
		&
		\mathcal{M}_{\sigma s}^2 + 2\omega\,\partial_{\tilde{t}} \\[1ex]
		\mathcal{M}_{s\sigma}^2 - 2\omega\,\partial_{\tilde{t}}
		& k^2/a^2 + \mathcal{M}_{ss}^2 + 3\omega^2
	\end{pmatrix}.
\end{equation}

Within the slow-roll regime, the coefficients vary only weakly on sub-horizon scales, allowing a local diagonalization of \(\mathbf{\Omega}^2_{\mathbf{k}}\) at each instant \(\tilde{t}\). Denoting the complex eigenvalues by
\begin{equation}
	\lambda_\pm^2(k,\tilde{t}) = \omega_\pm^2(k,\tilde{t})
	- i\,\omega_\pm(k,\tilde{t})\,\Gamma_\pm(k,\tilde{t}),
	\qquad
	\omega_\pm^2,\,\Gamma_\pm \in \mathbb{R},
	\label{eq:eigenvalues-complex}
\end{equation}
then $\Gamma_\pm(k,\tilde{t})$ has units of frequency. Defining the instantaneous eigenmodes \(Q_\pm\) via the complex linear transformation \(\mathbf{Q} = \mathbf{S}(\tilde{t})\,\mathbf{Q}_{\rm eig}\), with \(\mathbf{Q}_{\rm eig} = (Q_+,Q_-)^T\),  and neglecting time derivatives of \(\mathbf{S}\) in a WKB-like approximation, the mode equations reduce to a damped harmonic oscillator with a complex frequency and width \(\Gamma_\pm\)
\begin{equation}
	\ddot{Q}_\pm + 3\tilde{H}\,\dot{Q}_\pm + \lambda_\pm^2(k,\tilde{t})\,Q_\pm \simeq 0,
	\label{eq:Qpm-eq}
\end{equation}
encoding the non-Hermitian nature of the inflaton sector, in direct analogy with the complex-mass scheme for unstable particles \cite{campos2026}.

\subsection{Approximate solution and damping factor}

Considering the deep interior of the horizon, \(k \gg a\tilde{H}\), the imaginary parts of \(\lambda_\pm^2\) are negligible, so we again impose the Bunch–Davies initial conditions for \(Q_\pm\) as in the Hermitian case. On super-horizon scales, \(k \ll a\tilde{H}\), and for slowly varying \(\omega_\pm,\Gamma_\pm\), an approximate solution of equation \eqref{eq:Qpm-eq} takes the form
\begin{equation}
	Q_\pm(k,\tilde{t}) \;\simeq\; Q_\pm^{\rm (H)}(k,\tilde{t})\,
	\exp\!\left[- \frac{1}{2} \int_{\tilde{t}_\star}^{\tilde{t}}  \Gamma_\pm(k,\tilde{t}')\,{\rm d}\tilde{t}'\right],
	\label{eq:Qpm-solution}
\end{equation}
where \(Q_\pm^{\rm (H)}\) denotes the corresponding solution obtained in the purely Hermitian limit \(\Gamma_\pm(k,\tilde{t}) \to 0\). 
The exponential parameter represents the dissipative transfer of energy from the inflaton perturbations to the effective reheating bath. We can formally rewrite the system in terms of the adiabatic mode, obtaining
\begin{equation}
	Q_\sigma(k,\tilde{t})=b_+(k,\tilde{t})\,Q_+(k,\tilde{t})
	+ b_-(k,\tilde{t})\,Q_-(k,\tilde{t}),
\end{equation}
where the complex coefficients \(b_\pm(k,\tilde{t})\) are determined by the eigenvectors of the matrix \(\mathbf{\Omega}^2_{\mathbf{k}}\), given by equation \eqref{eq:mat_eigenx}. In regimes where the adiabatic direction follows a single eigenmode (when the trajectory is nearly straight and the entropy mode is heavy), then \(Q_\sigma \simeq Q_+\) up to a
constant phase, allowing us to write 
\begin{equation}
	Q_\sigma(k,\tilde{t})
	\;\simeq\;
	Q_\sigma^{\rm (H)}(k,\tilde{t})\,
	\exp\!\left[
	- \frac{1}{2}
	\int_{\tilde{t}_\star}^{\tilde{t}}
	\Gamma_\zeta(k,\tilde{t}')\,{\rm d}\tilde{t}'
	\right],
	\label{eq:Qsigma-damped}
\end{equation}
where  \(\Gamma_\zeta(k,\tilde{t})\) is the effective decay rate for the
curvature mode, which is a suitable combination of \(\Gamma_\pm(k,\tilde{t})\) weighted by the adiabatic projection. In practice, \(\Gamma_\zeta(k,\tilde{t})\) is controlled by the imaginary parts of the projected mass matrix \(\mathcal{M}^2_{AB}\) and related to the relevance parameter \(\mathcal{R}(N)\) \cite{campos2026}
\begin{equation}\label{eq:relevance}
	\mathcal{R}(N)=\frac{|\rho_I|}{\sqrt{\rho_R^{\,2}+\rho_I^{\,2}}}\in[0,1],
\end{equation}
where $\rho=\rho_R+i\rho_I$ is given by
\begin{equation}
	\rho_R = \frac{1}{2}\left(\dot\phi^{\,2}+\dot\chi^{\,2}\right)+V_R-3H\dot F,\quad\rho_I = V_I.
\end{equation}
and by the effective decay rate \(\Gamma_{\rm eff}(k,\tilde{t})\) that characterizes the background dynamics in the strongly non-Hermitian regime.
\begin{figure}[t!]
	\centering
	\includegraphics[width=0.8\linewidth]{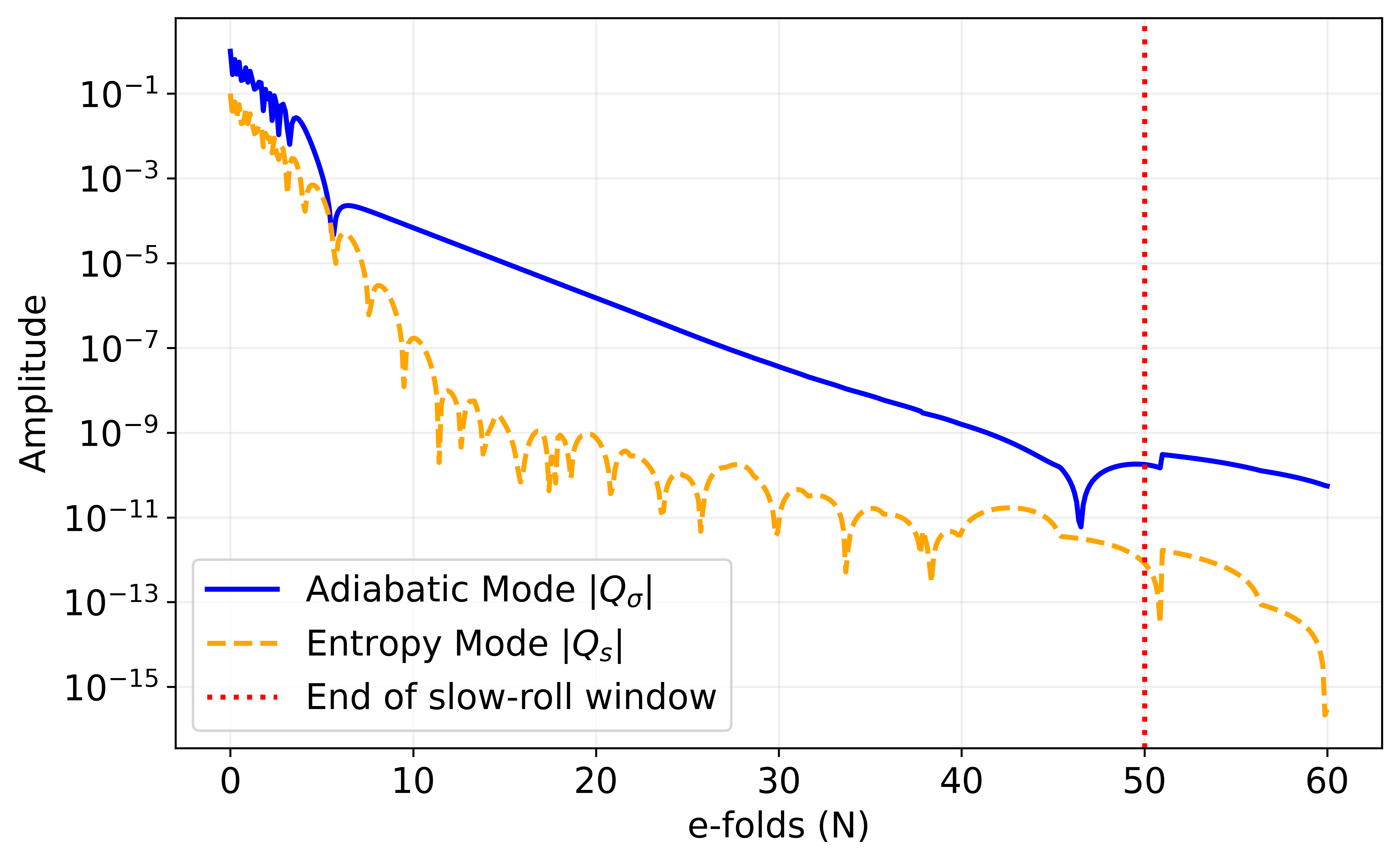}
	\caption{Numerical evolution of $|Q_{\sigma}|$ and $|Q_{s}|$ perturbation amplitudes versus e-folds $N$. The solid blue curve shows the adiabatic mode, and the dashed orange curve shows the strongly suppressed entropy mode. The vertical red dotted line at $N \approx 50$ marks the end of slow-roll inflation.}
	\label{fig:perturbation_evolution}
\end{figure}

Figure \ref{fig:perturbation_evolution} shows the numerical evolution of  $|Q_{\sigma}|$ and $|Q_{s}|$ for the CIF representation parameters. During the observable CMB window ($N \lesssim 50$), the adiabatic mode $|Q_{\sigma}|$ dominates and decays smoothly on super-horizon scales, while the entropy mode $|Q_{s}|$ is suppressed by several orders of magnitude, indicating that, in the quasi-Hermitian slow-roll regime, $V_I$ is a negligible perturbation, effectively decoupling isocurvature fluctuations. As a result, the effectively single-field $\alpha$-attractor mapping reliably predicts the spectral index $n_s$ and tensor-to-scalar ratio $r$, with an essentially vanishing isocurvature fraction $\beta_{iso}$.

Near the end of the slow-roll phase (red dotted line), the adiabatic evolution shows a sharp dip, $N \approx 47$, followed by a brief excitation, marking the activation of the non-Hermitian sector. As the inflaton trajectory begins to rotate in the internal field space $(x, \theta)$, the coupling between the real and imaginary parts of the potential becomes dynamically important. In the post-slow-roll regime ($N > 50$), the perturbations enter geometric reheating, where complex mass eigenvalues render the mode equations non-unitary and increase the damping of fluctuations. 

\subsection{Curvature power spectrum}

The comoving curvature perturbation is related to the adiabatic mode by equation \eqref{eq:zeta-from-Qsigma}, so the dimensionless power spectrum given by equation~\eqref{eq:power_spectra} at \(\tilde{t}\) can be written as
\begin{equation}\label{eq:power_spectrum_2}
	P_\zeta(k,\tilde{t}) = \frac{k^3}{2\pi^2} \left\langle\left|\zeta(k,\tilde{t})\right|^2\right\rangle,
\end{equation}
and using \eqref{eq:Qsigma-damped}, we obtain
\begin{equation}
	P_\zeta(k,\tilde{t})
	\;\simeq\;
	P_\zeta^{\rm (H)}(k,\tilde{t})\,
	\exp\!\left[
	- \int_{\tilde{t}_\star}^{\tilde{t}}
	\Gamma_\zeta(k,\tilde{t}')\,{\rm d}\tilde{t}'
	\right],
	\label{eq:Pzeta-damped}
\end{equation}
where \(P_\zeta^{\rm (H)}\) denotes the Hermitian prediction, given by the single-field \(\alpha\)-attractor prediction extended to the two-field Einstein-frame mapping, and the exponential factor accounts for the dissipative suppression (or enhancement, in principle) induced by the imaginary part of the complex potential. Observe that equation \eqref{eq:Pzeta-damped} summarizes how non-Hermiticity affects the curvature spectrum, showing that when $V_I$ is negligible during the observable slow-roll window, \(\Gamma_\zeta(k,\tilde{t})\) remains small for all CMB modes, and the correction factor is essentially unity. This agrees with the regime of Ref.~\cite{campos2026}, where \(\mathcal{R}(N_\star)\ll 1\) and \(\Delta N_H \ll 1\) hold, so the CMB observables \(n_s\) and \(r\) are indistinguishable from those of a purely Hermitian \(\alpha\)-attractor model. In contrast, modes that remain inside the horizon until the strongly non-Hermitian phase, such as small-scale scalar perturbations or reheating-era GWs, can experience substantial damping or spectral reshaping, offering potential phenomenological signatures of the CIF sector beyond the CMB window.
\begin{figure}[t]
	\centering
	\includegraphics[width=0.8\linewidth]{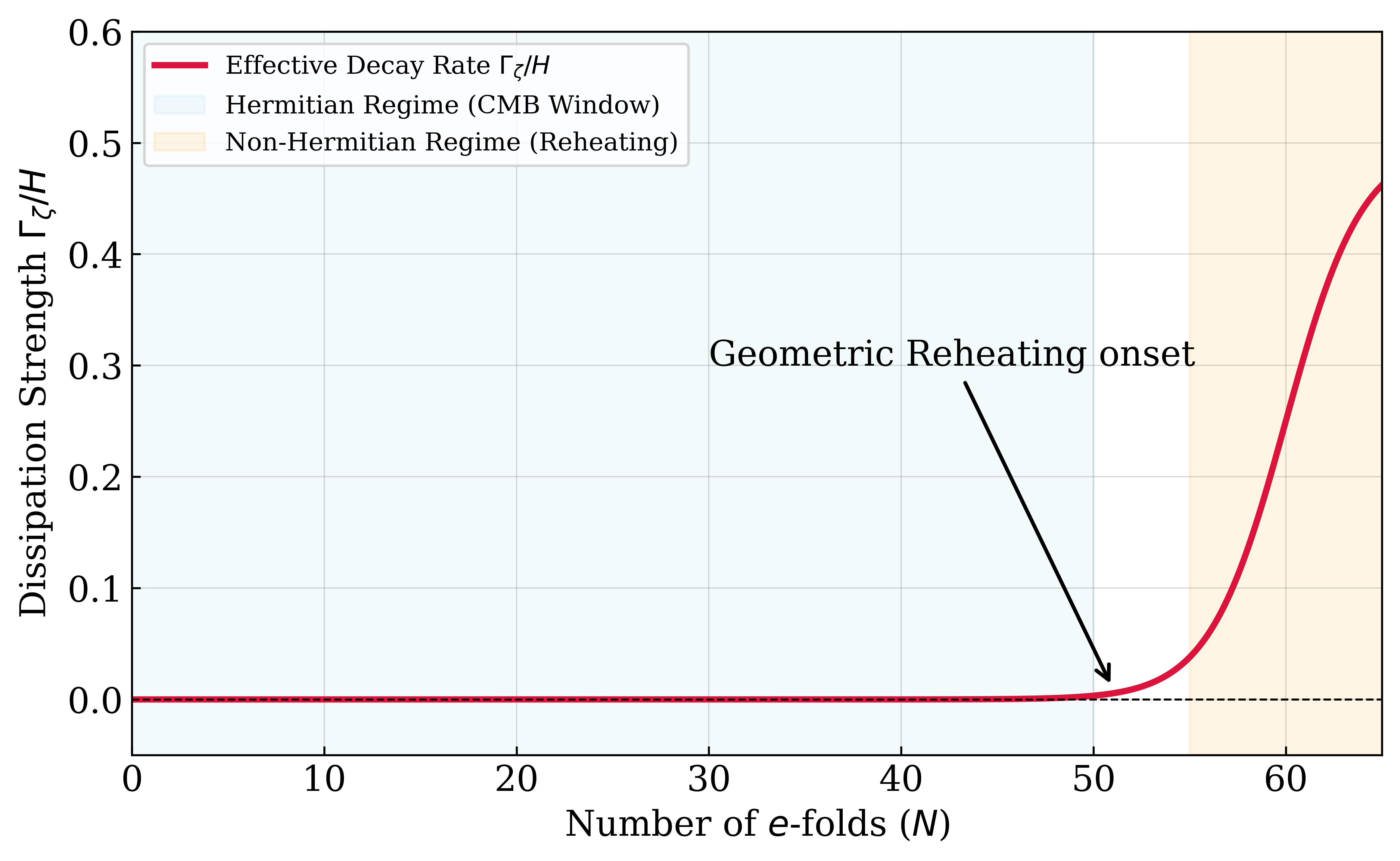}
	\caption{Activation of the non-Hermitian dissipation strength, $\Gamma_{\zeta}/H$, as a function of e-fold time $N$. The blue region ($N < 50$) denotes an effectively Hermitian regime in which the imaginary potential is negligible, preserving standard CMB predictions. The orange region ($N > 55$) marks the non-Hermitian reheating era, where the rapid growth of the decay rate signals activation of the dissipative sector as the inflaton trajectory rotates in internal field space, initiating geometric reheating.}
	\label{fig:dissipation_activation}
\end{figure}

\section{Phenomenological implications}
\label{sec:pheno}

Table \ref{tab:CMB-observables} shows that the non-Hermitian CIF model preserves the predictive power of standard $\alpha$-attractors, remains within the $1\sigma$ confidence region of the combined Planck and BK18 data, and simultaneously introduces new dynamics at the end of inflation.
\begin{table}[h!]
	\centering
	\caption{Comparison of CMB observables for the standard Hermitian $\alpha$-attractor, the proposed CIF model, and the latest observational constraints. The CIF values are evaluated in the early-time regime ($N \approx 50-60$), where non-Hermitian effects are subdominant.}
	\label{tab:CMB-observables}
	\begin{tabular}{@{}lccc@{}}
		\textbf{Observable} & \textbf{Hermitian $\alpha$-attractor} & \textbf{Non-Hermitian CIF} & \textbf{Observational Limits} \\ 
		\hline
		$n_s$ (Scalar Tilt) & $\approx 0.967$ & $0.967 + \mathcal{O}(\Gamma_\zeta/H)$ & $0.9649 \pm 0.0042$ \cite{Aghanim:2018eyx} \\
		$r$ (Tensor-to-scalar) & $\approx 0.003$ & $\approx 0.003$ & $< 0.036$ \cite{ade2021} \\
		$\alpha_s$ (Running) & $\approx -5 \times 10^{-4}$ & $\approx -5 \times 10^{-4}$ & $-0.0045 \pm 0.0067$ \cite{Aghanim:2018eyx} \\
		$f_{\rm NL}^{\rm local}$ & $\ll 1$ & $\ll 1$ & $-0.9 \pm 5.1$ \cite{Aghanim:2019bto} \\
		\hline
	\end{tabular}
\end{table}

Table \ref{tab:reheating-params} shows the impact of non-Hermitian dynamics on the post-inflationary epoch. Observe that by increasing the asymmetry parameter $\Delta\epsilon$, one raises the effective decay rate $\Gamma_{\rm eff}$ and shortens the reheating duration $N_{\rm reh}$, allowing high reheating temperatures without requiring very large microphysical couplings. 
\begin{table}[h]
	\centering
	\caption{The reheating duration $N_{\rm reh}$ and the final temperature $T_{\rm reh}$ are calculated for various non-minimal coupling strengths $\xi$ and asymmetry parameters $\Delta\epsilon$, assuming $m = 10^{-6}M_{\rm P}$ and $\lambda = 10^{-13}$.}
	\label{tab:reheating-params}
	\begin{tabular}{@{}ccccc@{}}
		\hline
		\textbf{Coupling $\xi$} & \textbf{Asymmetry $\Delta\epsilon$} & \textbf{Decay Rate $\Gamma_{\rm eff}/H$} & \textbf{Duration $N_{\rm reh}$} & \textbf{Temp. $T_{\rm reh}$ (GeV)} \\ 
		\hline
		10  & 0.1 & $1.2 \times 10^{-3}$ & $\approx 8.5$ & $\approx 10^{13}$ \\
		10  & 0.5 & $4.8 \times 10^{-2}$ & $\approx 4.2$ & $\approx 10^{14}$ \\
		\hline
		50  & 0.1 & $2.5 \times 10^{-3}$ & $\approx 6.1$ & $\approx 5 \times 10^{13}$ \\
		50  & 0.5 & $9.2 \times 10^{-2}$ & $\approx 2.8$ & $\approx 3 \times 10^{15}$ \\
		\hline
		100 & 0.8 & $2.1 \times 10^{-1}$ & $\approx 1.5$ & $\approx 10^{16}$ \\
		\hline
	\end{tabular}
\end{table}

As shown in Table \ref{tab:GW-observatories}, $V_I$ induces a scale-dependent suppression of the stochastic gravitational-wave background (SGWB) amplitude \cite{Maggiore2020,Kawamura2011}. Low-frequency observatories like the Laser Interferometer Space Antenna (LISA) are largely unaffected, while high-frequency probes such as the Einstein Telescope (ET) \cite{Maggiore2020}, the Big Bang Observer (BBO) \cite{Crowder2005}, and the Deci-hertz Interferometer Gravitational-wave Observatory (DECIGO) \cite{Kawamura2011} may be sensitive to the asymmetry parameter $\Delta\epsilon$, providing an observational test of non-unitary early-Universe dynamics, potentially distinguishing the CIF from Hermitian models.
\begin{table}[h]
	\centering
	\caption{Comparison of GW observatory sensitivities with the predicted SGWB from the CIF model. The damping effect from non-Hermiticity ($\Delta\epsilon > 0$) is most pronounced in the high-frequency ranges probed by BBO and ET.}
	\label{tab:GW-observatories}
	\begin{tabular}{@{}lcccc@{}}
		\hline
		\textbf{Observatory} & \textbf{Frequency Range} & \textbf{Sensitivity ($h^2\Omega_{\rm GW}$)} & \textbf{CIF ($\Delta\epsilon=0$)} & \textbf{CIF ($\Delta\epsilon=0.5$)} \\ 
		\hline
		LISA    & $10^{-4} - 10^{-1}$ Hz & $\approx 10^{-13}$ & $\approx 10^{-15}$ & $\approx 10^{-15}$ \\
		BBO     & $10^{-1} - 10^{1}$ Hz  & $\approx 10^{-17}$ & $\approx 10^{-16}$ & $\approx 10^{-17}$ \\
		DECIGO  & $10^{-1} - 10^{0}$ Hz  & $\approx 10^{-16}$ & $\approx 10^{-15}$ & $\approx 10^{-16.5}$ \\
		ET (Einstein Tel.) & $10^{0} - 10^{4}$ Hz   & $\approx 10^{-14}$ & $\approx 10^{-15}$ & $\approx 10^{-18}$ \\
		\hline
	\end{tabular}
\end{table}




\subsection{Scalar spectral tilt and running}
\label{subsec:scalar-running}


Defining the scalar tilt and running by \cite{Kosowsky:1995aa,Liddle:2000cg}
\begin{equation}\label{eq:tilt}
	n_s(k) - 1 \;\equiv\; \frac{{\rm d}\ln P_\zeta}{{\rm d}\ln k},
	\qquad
	\alpha_s(k) \;\equiv\; \frac{{\rm d}n_s}{{\rm d}\ln k},
\end{equation}
one writes
\begin{equation}
	\ln P_\zeta(k) =
	\ln P_\zeta^{\rm(H)}(k) - \mathcal{I}(k),
	\qquad
	\mathcal{I}(k)
	\equiv
	\int_{N_\star(k)}^{N_{\rm reh}}
	\frac{\Gamma_\zeta(k,N')}{H(N')}\,{\rm d}N',
\end{equation}
where \(k = a_\star H_\star\) and switched to e-fold time. Differentiating equation \eqref{eq:tilt} with respect to \(\ln k \simeq N_\star\)
gives
\begin{equation}
	n_s(k)-1
	\;=\;
	\bigl[n_s^{\rm(H)}(k)-1\bigr]
	- \frac{{\rm d}\mathcal{I}}{{\rm d}\ln k},
	\qquad
	\alpha_s(k)
	\;=\;
	\alpha_s^{\rm(H)}(k)
	- \frac{{\rm d}^2\mathcal{I}}{{\rm d}(\ln k)^2},
\end{equation}
with \(n_s^{\rm(H)}\) and \(\alpha_s^{\rm(H)}\) being the Hermitian
\(\alpha\)-attractor predictions. When \(k\)-dependence of \(\Gamma_\zeta(k,N)\) is negligible at fixed \(N\), the dominant scale dependence arises from the lower integration limit
\begin{equation}
	\frac{{\rm d}\mathcal{I}}{{\rm d}\ln k}
	\;\simeq\; - \frac{\Gamma_\zeta(N_\star)}{H(N_\star)},
\end{equation}
and thus, one obtains
\begin{equation}
	n_s(k) - 1
	\;\simeq\;
	\bigl[n_s^{\rm(H)}(k)-1\bigr]
	+ \frac{\Gamma_\zeta(N_\star)}{H(N_\star)},
	\label{eq:ns-delta-Gamma}
\end{equation}
\begin{equation}
	\alpha_s(k)
	\;\simeq\;
	\alpha_s^{\rm(H)}(k)
	+ \frac{1}{H_\star}
	\frac{{\rm d}}{{\rm d}N_\star}
	\left[
	\frac{\Gamma_\zeta(N_\star)}{H(N_\star)}
	\right].
	\label{eq:alphas-delta-Gamma}
\end{equation}

Equations \eqref{eq:ns-delta-Gamma} and \eqref{eq:alphas-delta-Gamma} show that $V_I$ corrects the Hermitian tilt and running by terms of order \(\Gamma_\zeta/H\) and its derivative. In the parameter space favored by CMB data, we found that \(\mathcal{R}(N_\star)\ll 1\) and \(\Delta N_H\ll 1\) at CMB e-folds, implying \(\Gamma_\zeta/H \ll 1\) in that range \cite{campos2026}. 

\begin{figure}[t]
	\centering
	\includegraphics[width=0.8\linewidth]{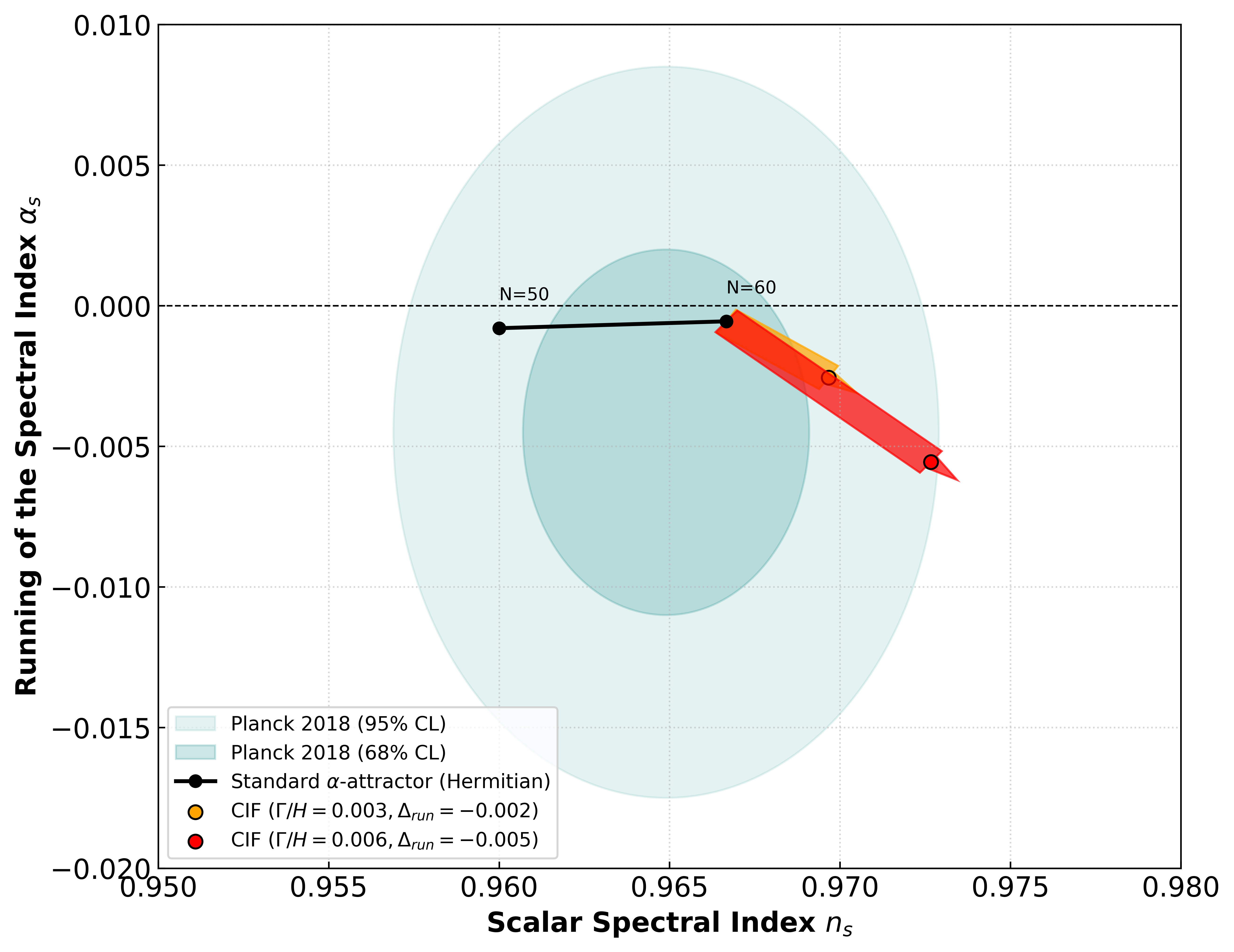}
	\caption{Sensitivity of the scalar spectral index $n_s$ and its running $\alpha_s$ to non-Hermitian dissipation. Light blue regions show the 68\% and 95\% CL Planck 2018 constraints. The solid black line is the standard Hermitian $\alpha$-attractor prediction for $N \in [50,60]$. Orange and red regions show the CIF-induced shift for increasing effective decay rate $\Gamma/H$ and its running correction $\Delta_{run}$. Small dissipation remains consistent with CMB data, whereas larger non-Hermitian effects push the model toward the edge of the observational contours.}
	\label{fig:ns_alphas_constraints}
\end{figure}

Figure \ref{fig:ns_alphas_constraints} shows the dependence of the non-Hermitian dynamics on the large-scale scalar observables by projecting CIF model predictions onto the $n_s - \alpha_s$ plane. 
For the standard Hermitian case (black solid line), the model is consistent with the Planck 2018 best-fit region with high accuracy, as expected for $\alpha$-attractor scenarios. Introducing a non-Hermitian sector shifts $n_s$ to higher values and makes the running $\alpha_s$ more negative. The orange and red wedges show this, corresponding to dissipation regimes $\Gamma/H = 0.003$ and $0.006$, respectively.

Our numerical analysis shows that, in the CMB-favored region where $\Gamma_{\zeta}/H \ll 1$ during the observable window, the model is indistinguishable from the Hermitian limit. However, the precise measurements of $\alpha_s$ can strongly constrain the onset of geometric reheating. If dissipation becomes significant too close to the end of slow roll, the resulting changes in the scalar tilt and running would push the model outside the $95\%$ CL Planck contours. Thus, requiring consistency with large-scale observations places an upper bound on the non-Hermitian deformation $\Delta \epsilon$ and dissipation rate $\Gamma_{\zeta}$, so dissipative effects mainly appear at smaller scales and higher frequencies, as in primordial GW damping \cite{Aghanim:2018eyx}.



\begin{figure}[t!]
	\centering
	\includegraphics[width=0.8\linewidth]{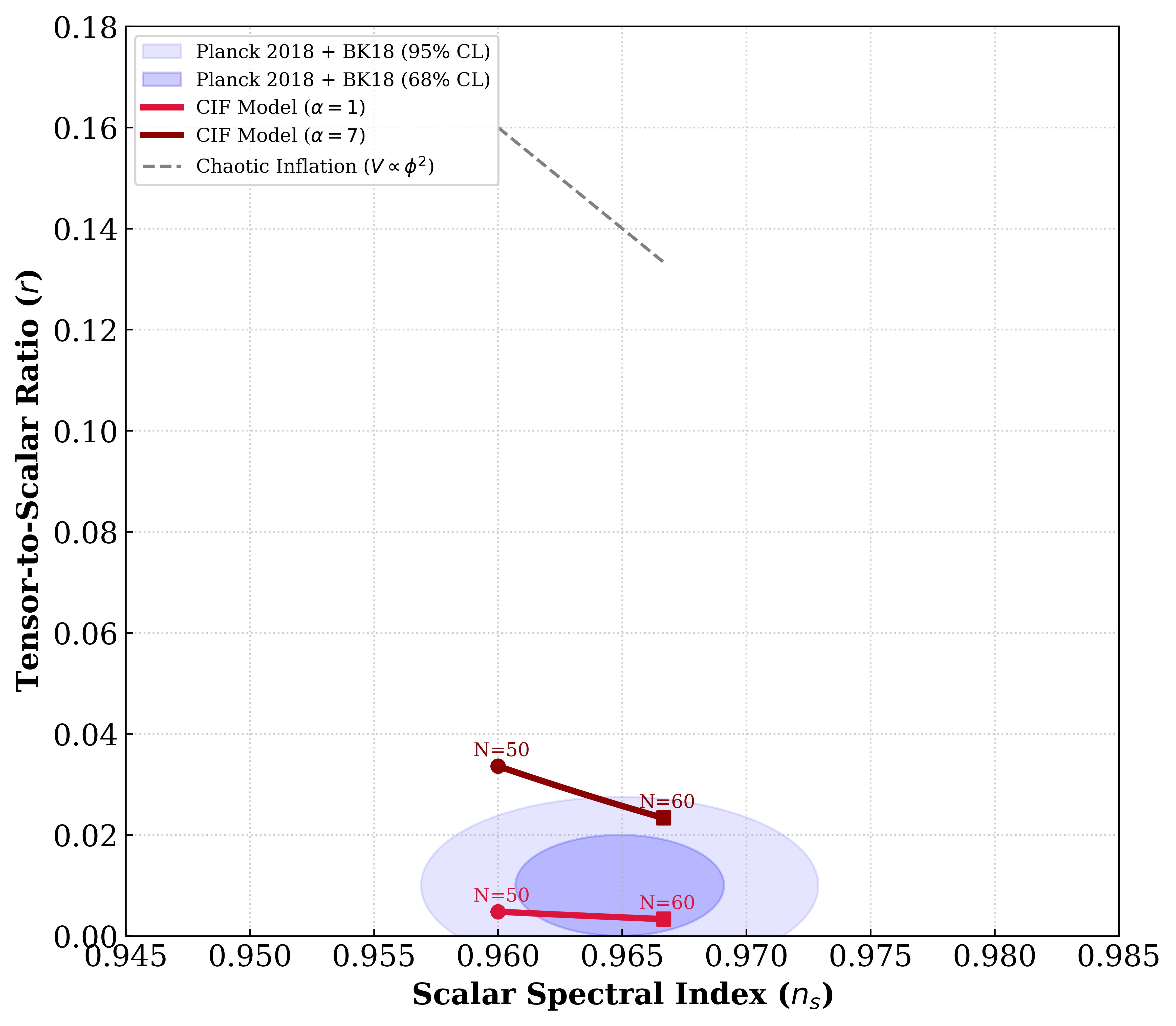}
	\caption{Predictions of the CIF model in the $n_s - r$ plane compared with Planck 2018 and BK18 constraints. The shaded blue ellipses show the 68\% and 95\% CL regions. The solid red and dark red lines correspond to the CIF model for $\alpha=1$ and $\alpha=7$, respectively, with $N \in [50, 60]$ e-folds. The grey dashed line denotes chaotic inflation ($V \propto \phi^2$), which is now excluded by the data. The CIF model fits the experimental contours well and suppresses the tensor-to-scalar ratio $r$ through its non-minimal coupling.}
	\label{fig:ns_r_plane}
\end{figure}


\subsection{Gravitational waves from geometric reheating}
\label{subsec:GW-pheno}

Of course, here $V_I$ does not contribute to the tensor sector at quadratic order, but it does change the background equation of state (EOS) and the effective decay history during geometric reheating, modifying the transfer function of any GW background generated before or during this epoch. The evolution of a tensor mode $h_k$ in a flat FRW background is governed by the linearized Einstein equations \cite{Maggiore:2018sht, Caprini:2018mtu}
\begin{equation}\label{eq:tensor_equation_1}
	h_k'' + 2H h_k' + k^2 h_k = \Pi_k,
\end{equation}
where primes denote derivatives with respect to \(\eta\) and \(\Pi_k\) is a source term (relevant during preheating or phase transitions). Moreover, $\Omega_{\rm GW}(k)$ is defined as \cite{Maggiore:2018sht}
\begin{equation}
	\Omega_{\rm GW}(k) 
	\;\equiv\; 
	\frac{1}{\rho_c} \frac{d\rho_{\rm GW}}{d\ln k},
\end{equation}
where $\rho_c = 3H_0^2 M_{\rm P}^2/(8\pi)$ is the present critical density, and $\rho_{\rm GW}$ is the GW energy density. This dimensionless spectrum gives the fractional GW energy density at scale $k$ and is the main observable for GW detectors searching for SGWB. In terms of $h_k(t)$, one writes \cite{Maggiore:2018sht}
\begin{equation}
	\Omega_{\rm GW}(k,t) 
	\;=\; 
	\frac{1}{24} \left(\frac{k}{aH}\right)^2 \overline{|h_k(t)|^2},
\end{equation}
where the overline denotes averaging over polarization states and 
propagation directions. For primordial GWs generated during inflation, the present-day spectrum is given by
\begin{equation}\label{eq:primordial_spectrum}
	\Omega_{\rm GW}(k,t_0) 
	\;=\; 
	\Omega_{\rm GW}^{\rm(prim)}(k)  \, T_{\rm GW}^2(k),
\end{equation}
where $\Omega_{\rm GW}^{\rm(prim)}(k)$ is the primordial spectrum at the end of inflation, $t_0$ denotes the present epoch, and \(T_{\rm GW}(k)\) is a transfer function determined by the background expansion history and any additional damping mechanisms, such as anisotropic stress or dissipative effects. Moreover, $T_{\rm GW}^2(k)$ represents the subsequent evolution through radiation domination, matter domination, and dark energy domination.



The present-day GW energy spectrum can be written from \eqref{eq:primordial_spectrum} as 
\begin{equation}
	\Omega_{\rm GW}(k,0)
	\;\simeq\;
	\Omega_{\rm GW}^{\rm(prod)}(k)\,   T_{\rm GW}^2(k),
\end{equation}
where \(\Omega_{\rm GW}^{\rm(prod)}(k)\) encodes the amplitude at $t_0=0$. In the present framework, $V_I$ modifies \(\mathcal{R}(N)\) and the real part of \(\omega(N)\), which together define an effective fluid that interpolates between quasi–de Sitter inflation and radiation domination.

Modes that re-enter the horizon during the geometric reheating phase experience a non-standard background characterized by \(w_{\rm eff}(N)\neq 1/3\), along with a non-negligible fraction of the energy density residing in the non-Hermitian sector when \(\mathcal{R}(N)\sim \mathcal{O}(1)\), which modifies the GW transfer function by altering the mode redshifting and hence the degree of suppression or enhancement relative to the radiation-dominated case. To show the leading impact of this sector on the GW transfer function, it is convenient to introduce a phenomenological
damping factor \(\mathcal{D}_{\rm GW}(k)\) 
\begin{equation}\label{eq:damping_factor}
	\Omega_{\rm GW}(k,0)
	\;\simeq\;
	\Omega_{\rm GW}^{\rm(std)}(k,0)\,
	\mathcal{D}_{\rm GW}(k),
\end{equation}
where \(\Omega_{\rm GW}^{\rm(std)}\) is the spectrum in a standard reheating scenario with the same background potential but a purely Hermitian inflaton, and \(\mathcal{D}_{\rm GW}(k)\) summarizes the additional damping (or enhancement) induced by $V_I$. A simple parameterization, inspired by the damping of primordial GWs by free-streaming neutrinos \cite{Weinberg:2003ur}, is given by
\begin{equation}
	\mathcal{D}_{\rm GW}(k)
	\;\simeq\;
	\exp\!\left[
	- \int_{N_{\rm in}(k)}^{N_{\rm reh}}
	\frac{f_{\rm NH}(N)}{1 + w_{\rm eff}(N)}\,{\rm d}N
	\right],
\end{equation}
where \(N_{\rm in}(k)\) is the e-fold number at which the mode \(k\) re-enters the horizon, \(w_{\rm eff}(N)\) is the effective real part of the EOS during geometric reheating, and \(f_{\rm NH}(N)\) is proportional to \(\mathcal{R}(N)\) that measures the strength of non-Hermitian effects in the tensor sector. Of course, in the Hermitian limit \(\mathcal{R}(N)\to 0\), \(f_{\rm NH}(N)\to 0\) and \(\mathcal{D}_{\rm GW}(k)\to 1\), recovering the standard result.



\subsection{Fermionic reheating and the effective equation of state}

To illustrate the calculation of $\mathcal{D}_{\rm GW}(k)$ in a concrete system, we consider perturbative reheating through the Yukawa interaction $\mathcal{L}_{\rm int} = - y\,\phi\,\bar\psi\psi$ between the real component of the CIF and a Dirac fermion. In the oscillatory regime, this gives an inflaton decay rate
$\Gamma_\phi \simeq y^2 m_\phi/(8\pi)$, which we identify with the
macroscopic effective decay rate $\Gamma_{\rm eff}(N)$ extracted from the imaginary part of the complex mass eigenvalues and the relevance parameter $\mathcal{R}(N)$ \cite{campos2026,Lozanov:2018zvx}.

The corresponding background evolution during reheating is governed by
\begin{align}
	\dot\rho_\phi &= -3H(1+w_\phi)\rho_\phi - \Gamma_\phi \rho_\phi, \\
	\dot\rho_R    &= -4H\rho_R + \Gamma_\phi \rho_\phi, \\
	H^2          &= \frac{\rho_\phi + \rho_R}{3M_{\rm P}^2},
\end{align}
where $w_\phi(N)$ is obtained from the CIF background solution and encodes the transition from quasi–de Sitter expansion to an oscillatory phase with $w_\phi\simeq 0$. Of course, the effective EOS controlling the tensor evolution is written as \cite{campos2026}
\begin{equation}
	w_{\rm eff}(N) = \frac{p_\phi + p_R}{\rho_\phi + \rho_R},
\end{equation}
where $p=p_R+ip_I$ with
\begin{equation}
	p_R = \frac{1}{2}\left(\dot\phi^{\,2}+\dot\chi^{\,2}\right)-V_R+\ddot F+2H\dot F,\quad p_I = -V_I, 
\end{equation}
computed numerically for any given choice of the microscopic parameters $(y,m_\phi)$ and the non-Hermitian sector.

\subsection{Definition of the damping factor $\mathcal{D}_{\rm GW}(k)$}

For a given fermionic reheating model specified by $(y,m_\phi)$, we integrate the homogeneous tensor equation~\eqref{eq:tensor_equation_1} from the end of inflation to the onset of the standard radiation-dominated era, using the previously obtained background evolution $a(\eta)$ and $w_{\rm eff}(N)$. The resulting tensor transfer function is written as
\begin{equation}
	T_{\rm GW}(k) \equiv \frac{h_k(\eta_0)}{h_k^{\rm prim}},
\end{equation}
where $h_k^{\rm prim}$ is the primordial tensor amplitude, and $\eta_0$ is the conformal time today. This function is computed using the non-Hermitian CIF reheating dynamics, including $V_I$ and the effective decay rate $\Gamma_{\rm eff}(N)$. 
To isolate the effect of the non-Hermitian fermionic reheating dynamics, we define $\mathcal{D}_{\rm GW}(k)$ by normalizing it to a reference model with instantaneous reheating (or constant $w$)
\begin{eqnarray}\label{eq:damping_factor_gw}
	\mathcal{D}_{\rm GW}(k) \equiv
	\frac{\Omega_{\rm GW}(k,0)}{\Omega_{\rm GW}^{\rm (std)}(k,0)} =
	\frac{T_{\rm GW}^2(k)}{T_{\rm GW,std}^2(k)},
\end{eqnarray}
where $T_{\rm GW,std}^2(k)$ serves as a baseline that represents what the tensor in a Hermitian model with the same inflationary potential but without the complex non-Hermitian effects during reheating. 


On large scales that remain super-horizon until matter-radiation equality ($k \ll k_{\rm eq}$, where $k_{\rm eq} \sim 0.01\,{\rm Mpc}^{-1}$ is the comoving scale entering the horizon at equality), the transfer function is nearly constant, $T_{\rm GW,std}^2(k) \simeq 1$. On smaller scales ($k \gg k_{\rm eq}$) that re-enter the horizon during the radiation-dominated era, the GW amplitude is suppressed relative to its primordial value. This evolution is encoded by $T_{\rm GW,std}^2(k)$, which shows a characteristic scale dependence \cite{Turner:1993vb, Boyle:2008se}
\begin{equation}\label{eq:stand_1}
	T_{\rm GW,std}^2(k) 
	\;\simeq\; 
	\left(\frac{k}{k_{\rm eq}}\right)^{-2} 
	\log^2\left(\frac{k}{k_{\rm eq}}\right),
	\qquad k \gg k_{\rm eq}.
\end{equation}

At even higher frequencies ($k \gtrsim k_{\nu} \sim 10^{-3}\,{\rm Mpc}^{-1}$),  neutrino free-streaming introduces additional exponential damping \cite{Weinberg:2003ur,Saikawa:2018rcs}
\begin{eqnarray}\label{eq:stand_2}
	T_{\rm GW,std}^2(k) \propto \exp[-2(k/k_{\nu})^2].
\end{eqnarray}

A convenient analytic approximation that interpolates between regimes shown by equations \eqref{eq:stand_1} and \eqref{eq:stand_2} is given by~\cite{Clarke:2020bil}
\begin{equation}
	T_{\rm GW,std}^2(k) 
	\;\simeq\; 
	\frac{1 + (k/k_{\rm eq})^{2\alpha}}{\bigl[1 + (k/k_{\rm eq})^2\bigr]^{\beta}} 
	\times 
	\exp\!\left[-2\left(\frac{k}{k_{\nu}}\right)^2\right],
\end{equation}
with empirically fitted exponents $\alpha \approx 0.67$ and $\beta \approx 1.17$.  
\begin{figure}[t]
	\centering
	\includegraphics[width=0.8\linewidth]{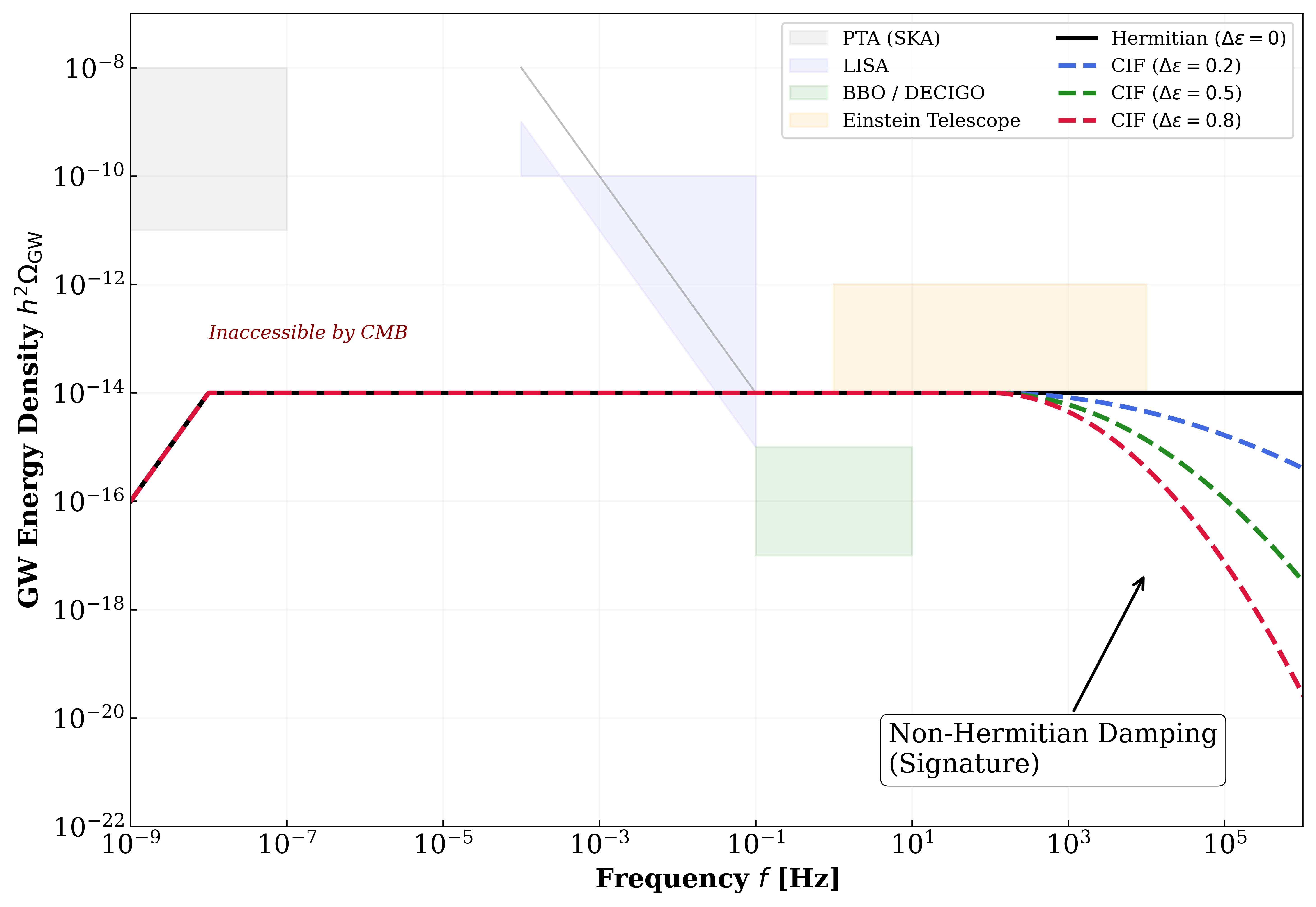}
	\caption{Predicted SGWB energy density for the Hermitian case ($\Delta\epsilon = 0$) and non-Hermitian asymmetries ($\Delta\epsilon = 0.2, 0.5, 0.8$). Sensitivity curves for PTA(SKA), LISA, BBO/DECIGO, and the ET are shown as shaded regions. At low frequencies, all models converge to the standard prediction ("Inaccessible by CMB"). At higher frequencies, the non-Hermitian sector produces a characteristic damping from dissipative energy transfer during geometric reheating, potentially observable by future GW interferometers.}
	\label{fig:sgwb_damping}
\end{figure}

As shown in Figure \ref{fig:sgwb_damping}, the CIF model predictions for $h^2\Omega_{GW}$ are identical to the Hermitian case at CMB scales, preserving observational consistency. A clear separation appears, however, at frequencies $f \gtrsim 10^2$ Hz, where the non-Hermitian damping signature falls within the maximum-sensitivity windows of the BBO and ET, providing a concrete observational target for detecting dissipative reheating dynamics. At low frequencies ($f \lesssim 10^{-7}$ Hz), relevant for the CMB, PTA, and Sky Kilometre Array (SKA) scales, the imaginary sector is negligible, and the background expansion is governed by $V_R$. The non-Hermitian consistency condition, $\mathcal{R}(N) \ll 1$, ensures that all models yield the same observables, explaining the "Inaccessible by CMB" label: standard cosmological probes may not distinguish between Hermitian and non-Hermitian inflation.

In the low-to-mid frequency range ($10^{-7} \text{ Hz} < f < 10^{1} \text{ Hz}$), the complex potential remains Hermitian, implying the predictions for the PTA(SKA) and LISA bands are identical to those of the $\alpha$-attractor models. This preservation of the large-scale spectrum is consistent with the freeze-out of adiabatic modes discussed in Section \ref{sec:perturbations}, ensuring that the CIF model does not violate current cosmological constraints. 

The "Non-Hermitian Deviation" appears at high frequencies, especially after the modes re-enter the horizon during the geometric reheating epoch. As the inflaton trajectory rotates in internal field space, the imaginary part of the mass matrix introduces $\mathcal{D}_{GW}(k)$, suppressing the tensor power spectrum, highly sensitive to the asymmetry parameter $\Delta\epsilon$: for $\Delta\epsilon = 0.8$ (red dashed line in Figure \ref{fig:sgwb_damping}), where we can see that the damping reaches several orders of magnitude in the high-frequency tail ($f > 10^4$ Hz).


\section{Final Remarks and Prospects}
\label{sec:conclusions}

We develop a perturbative framework for inflationary models featuring a CIF with non-minimal coupling and a non-Hermitian potential. By mapping the early-time dynamics onto an effectively conservative two-field model in the Einstein frame, we show that the CIF framework preserves the predictive power of the $\alpha$-attractor universality class while introducing geometric reheating as a novel mechanism to end inflation. These observational signatures provide a direct window into the underlying field-space geometry that motivates our central theoretical construction.

Across a broad range of the asymmetry parameter $\Delta\epsilon$, our numerical results in the $n_s$--$r$ and $n_s$--$\alpha_s$ planes remain consistent with Planck 2018 and BK18 constraints. Curvature perturbations freeze after horizon crossing, while isocurvature modes and local non-Gaussianities are strongly suppressed within the observable window, ensuring that the non-Hermitian deformation preserves the model's large-scale CMB predictions.

Near the end of slow roll, the inflaton trajectory turns, activating the dissipative ratio $\Gamma_{\zeta}/H$. This triggers geometric reheating and transfers energy to a radiation bath through non-unitary mode evolution. The resulting scale-dependent damping leaves a characteristic imprint on the SGWB: the low-frequency spectrum remains close to the Hermitian case, while frequencies above about $10^2\,\mathrm{Hz}$ are strongly suppressed. This provides a testable signature for future observatories such as the ET, BBO, and DECIGO.

Several extensions still remain to be explored. Although adding an imaginary potential provides a useful phenomenological approximation, a more rigorous derivation within the Schwinger-Keldysh (in-in) formalism for open quantum systems \cite{kamenev_book,Haehl2017} is needed to validate the dissipative operators and place the non-unitary dynamics on a firmer microscopic footing. Furthermore, fully tracking energy transfer requires specifying explicit couplings between the inflaton and Standard Model fields (e.g., via Yukawa-like or gauge interactions) to account for non-equilibrium thermalization \cite{campos2026} and to calculate the reheating temperature $T_{\text{rh}}$ from first principles. Additionally, parameter regimes with stronger trajectory bending at horizon crossing deserve further study, as they could enhance isocurvature modes and potentially address large-scale CMB anomalies.

Finally, because the imaginary sector $V_I$ remains only weakly constrained by current data, the degree of high-frequency suppression is subject to parametric uncertainty. Systematic scans of the $\Delta\epsilon$--$\xi$ parameter space using small-scale cosmological probes and high-frequency GW observatories will be essential to tightly bound or test these dissipative non-Hermitian mechanisms.

\section*{Acknowledgments}

SDC acknowledges the Federal University of São Carlos and the Applied Mathematics Laboratory for their institutional support.

\appendix

\section{Einstein-frame field-space metric}
\label{app:field_space_metric}

Here, we derive the field-space metric induced by the conformal transformation from the Jordan frame to the Einstein frame. We start from the Jordan-frame action for the two real components of the CIF given by definition \eqref{eq:def1} and
$|\Phi|^2 = \Phi\Phi^\ast = 1/2(\phi^2+\chi^2)$.

The Jordan-frame action is written as
\begin{equation}
	S_J = \int d^4x \sqrt{-g} \left[ \frac{1}{2}F(\phi,\chi)R-\frac{1}{2}\delta_{IJ}g^{\mu\nu}\partial_\mu \varphi^I \partial_\nu \varphi^J-V(\phi,\chi)\right],
	\label{eq:Jordan_action_appendix}
\end{equation}
where
\begin{equation}
	\varphi^I = (\phi,\chi),
	\qquad
	I,J=1,2,
\end{equation}
and
\begin{equation}
	F(\phi,\chi)=M_P^2-\xi(\phi^2+\chi^2)=M_P^2-2\xi|\Phi|^2.
	\label{eq:F_appendix}
\end{equation}

The condition $F(\phi,\chi)>0$ ensures that the conformal transformation is well defined and that the effective Planck mass remains positive. Performing the Weyl rescaling
\begin{equation}
	\tilde{g}_{\mu\nu}=\Omega^2(\phi,\chi)g_{\mu\nu},
	\qquad
	\Omega^2(\phi,\chi) =\frac{F(\phi,\chi)}{M_P^2},
	\label{eq:conformal_appendix}
\end{equation}
results in
\begin{equation}
	\sqrt{-g}=\Omega^{-4}\sqrt{-\tilde{g}},
	\qquad
	g^{\mu\nu} =\Omega^{2}\tilde{g}^{\mu\nu},
\end{equation}
and the Ricci scalar transforms as
\begin{equation}
	R=\Omega^2\left[\tilde{R}+6\tilde{\Box}\ln\Omega-6\tilde{g}^{\mu\nu}\partial_\mu\ln\Omega
	\partial_\nu\ln\Omega\right].
	\label{eq:Ricci_transform_appendix}
\end{equation}

The total derivative term is proportional to
\(\tilde{\Box}\ln\Omega\) can be discarded after integration by parts, assuming standard boundary conditions. The gravitational part of the action then becomes
\begin{equation}
	\sqrt{-g}\frac{1}{2}F R=\sqrt{-\tilde{g}}\left[
	\frac{M_P^2}{2}\tilde{R}-3M_P^2\tilde{g}^{\mu\nu}\partial_\mu\ln\Omega
	\partial_\nu\ln\Omega\right].
\end{equation}

Since
\begin{equation}
	\ln\Omega=
	\frac{1}{2}\ln\left(\frac{F}{M_P^2}\right),
\end{equation}
one has
\begin{equation}
	\partial_\mu\ln\Omega =
	\frac{1}{2F}F_{,I}\partial_\mu\varphi^I,
\end{equation}
where $F_{,I}\equiv \partial F/\partial \varphi^I$. Therefore, the contribution from the Weyl transformation to the scalar
kinetic sector is
\begin{equation}\label{eq:weyl}
	-3M_P^2 \tilde{g}^{\mu\nu} \partial_\mu\ln\Omega \partial_\nu\ln\Omega = -\frac{3M_P^2}{4F^2} F_{,I}F_{,J} \tilde{g}^{\mu\nu} \partial_\mu\varphi^I \partial_\nu\varphi^J.
\end{equation}

The original Jordan-frame kinetic term transforms as
\begin{equation}\label{eq:kine_jordan}
	-\frac{1}{2}\sqrt{-g}\, \delta_{IJ}g^{\mu\nu} \partial_\mu\varphi^I \partial_\nu\varphi^J =
	-\frac{1}{2}\sqrt{-\tilde{g}}\, \frac{M_P^2}{F} \delta_{IJ} \tilde{g}^{\mu\nu} \partial_\mu\varphi^I \partial_\nu\varphi^J.
\end{equation}

Combining contributions \eqref{eq:weyl} and \eqref{eq:kine_jordan}, the Einstein-frame action takes the form
\begin{equation}
	S_E =  \int d^4x\sqrt{-\tilde{g}} \left[ \frac{M_P^2}{2}\tilde{R}  -\frac{1}{2} \mathcal{G}_{IJ}(\phi,\chi) \tilde{g}^{\mu\nu} \partial_\mu\varphi^I \partial_\nu\varphi^J -U(\phi,\chi)
	\right],
	\label{eq:Einstein_action_appendix}
\end{equation}
where the Einstein-frame potential is given by equation \eqref{eq:einstein_pot},
and the field-space metric is written as
\begin{equation}
	\mathcal{G}_{IJ}(\phi,\chi) = \frac{M_P^2}{F(\phi,\chi)}\delta_{IJ} + \frac{3M_P^2}{2F^2(\phi,\chi)}  F_{,I}F_{,J}.
	\label{eq:general_field_metric_appendix}
\end{equation}

For the specific non-minimal coupling given by equation~\eqref{eq:ftheta},  
we have $F_{,\phi}=-2\xi\phi$ and $F_{,\chi}=-2\xi\chi$. Substituting these derivatives into equation~\eqref{eq:general_field_metric_appendix}, one obtains
\begin{equation}\label{eq:Gphiphi_appendix}
	\mathcal{G}_{\phi\phi} = \frac{M_P^2}{F} + \frac{6M_P^2\xi^2\phi^2}{F^2},\quad
	\mathcal{G}_{\chi\chi} = \frac{M_P^2}{F} + \frac{6M_P^2\xi^2\chi^2}{F^2},\quad
	\mathcal{G}_{\phi\chi} = \mathcal{G}_{\chi\phi} = \frac{6M_P^2\xi^2\phi\chi}{F^2}.
\end{equation}

Thus, on the \((\phi,\chi)\) basis, the line element in field space is
\begin{equation}
	d\ell^2  = \mathcal{G}_{\phi\phi}d\phi^2 + 2\mathcal{G}_{\phi\chi}d\phi d\chi + \mathcal{G}_{\chi\chi}d\chi^2=\frac{M_P^2}{F} \left(d\phi^2+d\chi^2\right) + \frac{6M_P^2\xi^2}{F^2} \left(\phi d\phi+\chi d\chi\right)^2.
	\label{eq:line_element_cartesian_appendix}
\end{equation}

It is useful to rewrite the metric in polar variables,
\begin{equation}
	\phi = x\cos\theta,
	\qquad
	\chi = x\sin\theta,
	\qquad
	x^2=\phi^2+\chi^2=2|\Phi|^2,
\end{equation}
and, in these coordinates, one has
\begin{equation}
	d\phi^2+d\chi^2 = dx^2+x^2d\theta^2,\quad
	\phi d\phi+\chi d\chi = xdx.
\end{equation}

Therefore,
\begin{equation}
	d\ell^2 = \mathcal{G}_{xx}(x)dx^2 + \mathcal{G}_{\theta\theta}(x)d\theta^2,
	\label{eq:polar_metric_appendix}
\end{equation}
with
\begin{equation}
	\mathcal{G}_{xx}(x) = \frac{M_P^2}{F(x)} + \frac{6M_P^2\xi^2x^2}{F^2(x)},\quad
	\mathcal{G}_{\theta\theta}(x) = \frac{M_P^2}{F(x)}x^2,
	\label{eq:Gthetatheta_appendix}
\end{equation}
where, now, one writes
\begin{equation}
	F(x)=M_P^2-\xi x^2.
\end{equation}

This form explicitly shows that the conformal transformation generates different kinetic weights for radial and angular motion in field space. In terms of the CIF, the canonical flat kinetic term satisfies
\begin{equation}
	\partial_\mu\Phi\,\partial^\mu\Phi^\ast =
	\frac{1}{2}\left[\partial_\mu\phi\,\partial^\mu\phi
	+\partial_\mu\chi\,\partial^\mu\chi\right].
\end{equation}

However, after the conformal transformation, the induced metric contains the additional rank-one contribution proportional to
\(F_{,I}F_{,J}\). Therefore, the full two-field Einstein-frame kinetic sector is not, in general, described by a single scalar function multiplying
\(\partial_\mu\Phi\,\partial^\mu\Phi^\ast\). Instead, the exact kinetic sector is
\begin{equation}
	-\frac{1}{2} \mathcal{G}_{IJ} \tilde{g}^{\mu\nu} \partial_\mu\varphi^I \partial_\nu\varphi^J.
\end{equation}

A scalar coefficient
\(\mathcal{G}(|\Phi|^2)\) may be introduced only in a reduced description,
for example, along an approximately radial trajectory, for which
\(d\theta\simeq 0\). In that case,
\begin{equation}
	d\ell^2
	\simeq
	\mathcal{G}_{xx}(x)dx^2,
\end{equation}
and the effective radial kinetic coefficient is
\begin{equation}
	\mathcal{G}_{\rm rad}(|\Phi|^2) = \frac{M_P^2}{M_P^2-2\xi|\Phi|^2} + \frac{12M_P^2\xi^2|\Phi|^2} {\left(M_P^2-2\xi|\Phi|^2\right)^2}.
	\label{eq:Gradial_complex_appendix}
\end{equation}

This is the quantity that should be identified with
\(\mathcal{G}(|\Phi|^2)\) when the background evolution is effectively radial. For a genuinely two-field trajectory with non-negligible angular motion, the full metric tensor
\(\mathcal{G}_{IJ}\) in equations \eqref{eq:Gphiphi_appendix} must be retained.

\end{document}